\documentclass{ws-mpla}
\usepackage[super]{cite}
\usepackage{graphicx}
\usepackage{xcolor}

\begin{document}

\markboth{W.-S. Hou and T. Modak}
{Probing TCNH Couplings at Colliders}

\catchline{}{}{}{}{}

\title{Probing Top Changing Neutral Higgs Couplings at Colliders
}

\author{\footnotesize Wei-Shu Hou
}

\address{Department of Physics, National Taiwan University, Taipei 10617,
Taiwan
\\
wshou@phys.ntu.edu.tw}

\author{Tanmoy Modak}

\address{Institut f\" ur Theoretische Physik,
Universit\" at Heidelberg,
69120 Heidelberg, Germany
\\
tanmoyy@thphys.uni-heidelberg.de}

\maketitle

\pub{Received (Day Month Year)}{Revised (Day Month Year)}

\begin{abstract}
The $h(125)$ boson, discovered only in 2012, is lower than the top quark in mass,
hence $t \to ch$ search commenced immediately thereafter, with current limits 
at the per mille level and improving.
As the $t \to ch$ rate vanishes with the $h$-$H$ mixing angle $\cos\gamma \to 0$,
we briefly review the collider probes of the top changing $tcH/tcA$ coupling 
$\rho_{tc}$ of the exotic $CP$-even/odd Higgs bosons $H/A$. 
Together with an extra top conserving $ttH/ttA$ coupling $\rho_{tt}$, 
one has an enhanced $cbH^+$ coupling alongside the familiar $tbH^+$ coupling, 
where $H^+$ is the charged Higgs boson.
The main processes we advocate are $cg \to tH/A \to tt\bar c,\; tt\bar t$ 
(same-sign top and triple-top), and $cg \to bH^+ \to bt\bar b$.
We also discuss some related processes such as $cg \to thh$, $tZH$
that depend on $\cos\gamma$ being nonzero, comment briefly 
on $gg \to H/A \to t\bar t, t\bar c$ resonant production,
and touch upon the $\rho_{tu}$ coupling.
\end{abstract}

\ccode{PACS Nos.: .}

\section{Introduction}	

It was proposed in the early 1990s, before the top quark was even discovered, 
that\cite{Hou:1991un} ``{\it $t \to ch$ or $h \to t\bar c$ decays 
could be quite prominent, if not the dominant, decay
modes of the top quark and some neutral Higgs bosons}'' ($h$).
The generic neutral Higgs boson $h$ (whether scalar or pseudoscalar)
in the two Higgs doublet model (2HDM) context
can have $\bar q_i q_j h$ couplings proportional\cite{Cheng:1987rs} to 
$\sqrt{m_i m_j}$, thereby evade effectively the low energy
flavor changing neutral current (FCNC) constraints
that so worried Glashow and Weinberg.\cite{Glashow:1976nt}
As much as this Cheng-Sher ansatz\cite{Cheng:1987rs} 
helped lead the way to identify $tch$ as plausibly 
the largest flavor changing neutral Higgs (FCNH) coupling, 
it was pointed out\cite{Hou:1991un} that $\sqrt{m_i m_j}$ dependence 
need not be literal,\footnote{
Hall and Weinberg also discussed~\cite{Hall:1993ca} $t \to ch$
under approximate $U(1)$ flavor symmetries.
}
but just ``{\it reflect fermion  mass and mixing hierarchies}'', i.e. the {flavor} enigma.
This is distinct from the two usual 2HDMs, Models I and II, where
the ``Natural Flavor Conservation'' (NFC) condition\cite{Glashow:1976nt}
is enforced by a $Z_2$ symmetry that completely eliminates
the second set of Yukawa couplings.
The 2HDM with extra Yukawa couplings, i.e. without an {\it ad hoc} $Z_2$ symmetry, 
but controlled by the emergent and therefore ``{\it natural}\," fermion mass-mixing
hierarchies, was christened\cite{Hou:1991un} Model~III.
With the top quark nowhere to be seen at the time, 
it was even suggested\cite{Hou:1993mw} that $m_t < M_W$ could still be 
possible, with $t \to bW^*$ obscured by e.g. $t \to ch$ (or $bH^+$).
For a general review of 2HDMs, see Ref.~\refcite{Branco:2011iw}.

But the top quark was discovered via the $t \to bW$ channel in 1995 at the Tevatron,
while it took almost two more decades for the elusive $h(125)$ boson 
to be discovered at the Large Hadron Collider (LHC).
To set up notation, we note the remarkable {\it emergent} phenomenon 
from LHC Run~1 data, called ``alignment'': the $h(125)$ boson 
was found to resemble rather closely\cite{Khachatryan:2016vau} 
the Higgs boson of the Standard Model (SM).
In the 2HDM framework, we denote $h(125)$ as $h$, while 
$H/A$ and $H^+$ are the exotic $CP$-even/odd and charged scalars, respectively.

After the top quark discovery,\cite{PDG} in context of the push (in the 1990s!) 
for a 500~GeV $e^+e^-$ linear collider and other future facilities, 
top FCNH was studied for 
$e^+e^-$,\cite{Atwood:1995ud, Hou:1995qh, BarShalom:1997tm, Hou:1997rr, BarShalom:1997sj, BarShalom:1999iy, Han:2001ap}
$\gamma\gamma$,\cite{Jiang:1998gm} 
and even\cite{Atwood:1995ej} $\mu^+\mu^-$ colliders.
One novel idea was that
$e^+e^- \to Z^* \to hA$ or $HA$ pair production can\cite{Hou:1995qh} 
give rise to $tt\bar c\bar c$ final states,\footnote{
 The conjugate process is always implied throughout the paper.
} 
i.e. same-sign top production.
Exotic scalar production by vector boson fusion with ``neutrino tags'',
i.e. the $e^+e^- \to t\bar c \nu_e\bar\nu_e$ process
was also proposed,\cite{BarShalom:1997tm, Hou:1997rr, BarShalom:1997sj} 
but this channel may not be promising, as the exotic scalars 
probably do not couple much with vector bosons due to alignment.
The same-sign top idea, however, was carried over to the LHC,
with the {\it direct} production process\cite{Hou:1997pm} of 
$cg \to tA$ followed by $A \to t\bar c$ decay, or $cg \to tt\bar c$. 
We will see that this process is still relevant, but we are getting ahead of ourselves.

Let us come back to the $t\to ch$ decay process.
In SM, $t\to ch$ can only arise at the loop level\footnote{
 We do not discuss loop-induced $tcZ$ and $tc\gamma$ couplings
 as they are less promising in terms of rate.
 For an early exposition, we refer to Ref.~\refcite{Atwood:1996vj},
 and references therein. 
 A subsequent review can be found in Ref.~\refcite{AguilarSaavedra:2004wm},
 which covers different models and processes, including tree level effects.
}  
hence {\it very} suppressed, i.e. at~\cite{Eilam:1990zc,Mele:1998ag} 
${\cal B}(t \to ch) \sim 10^{-15}$ and even smaller for $t \to uh$.
This is definitely out of reach,\footnote{
 The much larger\cite{Han:2013sea}
 ${\cal B}(t \to bW^*h) \simeq 2 \times 10^{-9}$ in SM is still out of reach.
}
but it implies that any observation
would indicate beyond SM (BSM) {\it New Physics}.
Ref.~\refcite{Eilam:1990zc} also discussed 2HDM II,
but since the authors corrected a sign error for the SM result
after Ref.~\refcite{Mele:1998ag} appeared, we do not quote their result.
It was later found\cite{Bejar:2000ub, Arhrib:2005nx} in 2HDM II 
that ${\cal B}(t \to ch) \sim 10^{-5}$ (or higher) can be reached, 
but this relies on large $\tan\beta$ enhancement, 
as is the study\cite{Baum:2008qm} of loop effects 
from a ``2HDM for the top quark''.\cite{Das:1995df}
A similar level can be reached in SUSY,\cite{Guasch:1999jp} 
but it relies on flavor changing gluino couplings in a time
far before the advent of the LHC.
SUSY loop effects were of course followed up,\cite{Cao:2007dk}
but became diminished\cite{Cao:2014udj} after $h(125)$ discovery.
Other avenues explored are, to name a few,
$R$-parity violation,\cite{Eilam:2001dh}
quark singlets,\cite{AguilarSaavedra:2002kr, Gaitan:2004by}
warped extra dimensions,\cite{Azatov:2009na}
mirror fermions,\cite{Yang:2013lpa}
composite Higgs,\cite{Azatov:2014lha}
and aligned 2HDM,\cite{Abbas:2015cua}
where the latter three are post $h(125)$ discovery.
None of these truly reach above $10^{-5}$, hence all fall short of 
a theoretical sensitivity estimate\cite{AguilarSaavedra:2000aj} 
at $\sim 10^{-4}$ for the LHC.

In contrast to the above suppression of $t \to ch$ rates,
the expectation in 2HDM~III is wide open, as the process is 
tree level,\cite{Hou:1991un, Iltan:2001yt, Aranda:2009cd}
which has also been studied via effective field theory (EFT).\cite{CorderoCid:2004vi, Larios:2004mx, Fernandez:2009vr}
With the advent of the LHC, and with early indications of a light Higgs boson, 
a detailed study\cite{Kao:2011aa} advocated $t \to ch$ followed 
by $h \to b\bar b$ as the discovery mode.
By the time the paper was published, however, ATLAS and CMS
had discovered\cite{PDG} the $h(125)$ boson. 
It was then pointed out that ${\cal B}(t\to ch)$ at 
the percent level\cite{Chen:2013qta} was still possible, 
and the search began in earnest.
Partially stimulated by the CMS hint\cite{Khachatryan:2015kon}
for $h \to \tau\mu$ decay from 8~TeV data, variants of 2HDM~III were 
discussed,\cite{Chiang:2015cba, Crivellin:2015hha, Botella:2015hoa, Chiang:2017fjr}
but soon died out when the hint was not confirmed by 13~TeV data.\cite{PDG}
The experimental pursuits, however, continued.

There is actually a catch for $t \to ch$ search: the $tch$ coupling 
is suppressed by $h$--$H$ mixing, i.e. between the two $CP$-even bosons, 
hence nonobservation could be due to small mixing rather than small $tcH$ coupling.
In this Brief Review, we focus on probing top changing neutral Higgs (TCNH) 
couplings at colliders via {\it direct} production processes. 
Rather than 2HDM~III, we shall call the 2HDM 
with extra Yukawa couplings ``general 2HDM'' (g2HDM). 
%
In what follows, we give our theoretical framework in Sec.~2,
then discuss $t \to ch$ search in Sec.~3.
We turn to our main theme of direct production processes 
via TCNH coupling in Sec.~4, where Sec.~4.1 covers 
the $cg \to tH/A \to tt\bar c,\; tt\bar t$ processes, namely 
same-sign top with jet and the more exquisite triple-top,
and Sec.~4.2 covers $cg \to bH^+ \to bt\bar b$.
In Sec.~5 we offer some discussion and cover a miscellany of topics,
and present our summary and prospects in Sec.~6.

\section{Theoretical Framework}

It is useful to trace the theory development since the early days
of the $t\to ch$ proposal.\cite{Hou:1991un}
The Cheng-Sher treatment, while evoking the mass-mixing hierarchy
to evade low energy FCNC constraints, was somewhat vague regarding the
Higgs bosons. The scenario does remove the usual but {\it ad hoc} $Z_2$ symmetry
that enforces the NFC condition\cite{Glashow:1976nt} of Glashow and Weinberg, and
more systematic works clarified\cite{Davidson:2005cw} the situation, 
both for the Higgs potential, and the Yukawa couplings.

We write the most general $CP$-conserving two Higgs doublet (without $Z_2$)
potential in the Higgs basis as\cite{Hou:2017hiw, Davidson:2005cw}
\begin{align}
 &V(\Phi,\, \Phi')
= \mu_{11}^2 \Phi^2 +\mu_{22}^2 \Phi'^2
        - \left(\mu_{12}^2 \Phi^\dagger \Phi' + {\rm h.c.}\right)
 +\frac{\eta_1}{2}\Phi^4 + \frac{\eta_2}{2}\Phi'^4 \notag\\
   &\ + \eta_3\Phi^2\Phi'^2
   + \eta_4\Phi^\dagger \Phi'^2 
    +\left\{\frac{\eta_5^{}}{2}(\Phi^\dagger \Phi')^2
   + \left[\eta_6^{}\Phi^2 + \eta_7^{}
      \Phi'^2\right]\Phi^\dagger \Phi' + {\rm h.c.}\right\}.
 \label{eq:pote2}
\end{align}
where $\langle \Phi\rangle \neq 0$ by\footnote{
 We note that, without a $Z_2$ symmetry, the usual notion of
 $\tan\beta = v_1/v_2$ is unphysical.\cite{Davidson:2005cw}
 Note also that the $\eta_6$ and $\eta_7$ terms in Eq.~(\ref{eq:pote2})
 would be absent in 2HDMs with $Z_2$ symmetry.
} 
$\mu_{11}^2  = - \frac{1}{2}\eta_1 v^2$,
while $\langle \Phi'\rangle = 0$ hence $\mu_{22}^2 > 0$. 
A second minimization condition, $\mu_{12}^2 = \frac{1}{2}\eta_6 v^2$, 
removes the soft $\mu_{12}^2$ term,
reducing the total number of parameters to nine,\cite{Hou:2017hiw} 
with $\eta_6$ the sole parameter for $h$-$H$ mixing, namely
\begin{align}
  M_\textrm{even}^2 =
  \left[\begin{array}{cc}
    \eta_1^{}v^2 & \ \eta_6^{} v^2 \\
    \eta_6^{} v^2 & \ m_A^2 + \eta_5v^2\\
    \end{array}\right],
 \label{eq:Msq-even}
\end{align}
for the $CP$-even Higgs mass matrix, where
$m_A^2 = \mu_{22}^2 +\frac{1}{2}(\eta_3 +\eta_4^{} - \eta_{5}^{})v^2$.
%
%
As mentioned, the {\it emergent} ``alignment'' phenomenon 
from LHC Run~1, that \cite{Khachatryan:2016vau}
the $h$ boson resembles rather closely the Higgs boson of SM, 
implies  that the $h$--$H$ mixing angle $\cos\gamma$ (which we 
denote as $c_\gamma$, and similarly $s_\gamma \equiv \sin\gamma$) is rather small.
We note\cite{Hou:2017hiw} the approximate relation near alignment,
\begin{align}
c_\gamma \cong \frac{\eta_6 v^2}{m_H^2 - m_h^2},
 \label{eq:c_gam}
\end{align}
as $s_\gamma \rightarrow 1$ more rapidly than $c_\gamma \rightarrow 0$.
Thus, it is quite intuitive that 
$\Phi$ serves as the ``mass-giver'' with $\mu_{11}^2 < 0$, while 
$\Phi'$ is the exotic doublet with an inertial mass parameter $\mu_{22}^2 > 0$,
and the ``soft'' $\mu_{12}^2$ term is eliminated. This is in contrast to 
the usual 2HDM I and II with $Z_2$ symmetry, where
both $m_{11}^2,\, m_{22}^2 < 0$, with the $m^2_{12}$ term
playing the dual role of $h$--$H$ mixing {\it and} inertial mass.

If one now takes the known $v \cong 246$ GeV as the sole scale parameter,
then the traditional, elementary notion of ``naturalness'' dictates that
all dimensionless quantities in $V(\Phi,\, \Phi')$, 
namely all $\eta_i$s and $\mu_{22}^2/v^2$ in Eq.~(\ref{eq:pote2}), 
ought to be ${\cal O}(1)$. 
It was shown explicitly in Ref.~\refcite{Hou:2017hiw} (see Fig.~1 of the reference)
that this ``naturalness'' can be maintained without running into conflict with alignment, 
i.e. $c_\gamma$ in Eq.~(\ref{eq:c_gam}) can be small with large parameter space.
Although $\eta_1 < \eta_6$ seems needed, both can be ${\cal O}(1)$,
and the dynamical mixing parameter $\eta_6$ actually helps 
push $\eta_1 v^2$ down to the physical $m_h^2$ 
by level repulsion,\footnote{
 The same level repulsion due to $\eta_6$ pushes
 $m_A^2 + \eta_5 v^2$ up to $m_H^2$ and helps maintain alignment.
} 
hence $\mu_{11}^2/v^2$ is indeed also ${\cal O}(1)$! 
This is in contrast even with SM, where the usual $\sqrt{\mu^2} \simeq 87$~GeV 
is a factor $\sim 2\sqrt 2$ lower than $v$, which is borderline on being ``natural''.
Ref.~\refcite{Hou:2017hiw} then asserts that, on top of 
mass-mixing hierarchy suppression in general,\cite{Hou:1991un} 
the unforeseen emergent phenomenon of alignment, i.e. small $c_\gamma$, 
further suppresses the FCNH couplings of the lighter $h$.
Thus, the combined effect of mass-mixing hierarchy and alignment 
can replace the brutal NFC condition, and one should therefore really 
view the usual $Z_2$ symmetry for what it is: {\it ad hoc}.

Though not spelled out in great detail in Ref.~\refcite{Hou:2017hiw}, 
the naturalness argument of ${\cal O}(1)$ parameters in $V(\Phi,\, \Phi')$,
together with alignment, implies the mass range
\begin{align} 
m_H, m_A, m_{H^+} \in (300, 600)\;{\rm GeV},
 \label{eq:mrange}
\end{align}
which seems just right for the LHC to probe. 
The point is, at the upper mass range and beyond, 
either one has $\eta_i > {\cal O}(1)$ (strong couplings), 
or $\mu_{22}^2/v^2 > {\cal O}(1)$ (decoupling) would set in.
The former heads toward nonperturbativity, making estimates unreliable,
while the latter would damp dynamical effects of interest. 
For the lower mass range, given that $m_H^2$ is raised by level repulsion, 
for $m_H < 300$ GeV or so would bring $m_A^2 + \eta_5 v^2$
closer to $\eta_1 v^2$, and even a weak $\eta_6$
could more easily generate $c_\gamma$,
and alignment becomes harder to sustain.
It should be clear that the sub-TeV range of Eq.~(\ref{eq:mrange}) 
is just right for the LHC to probe in the next two decades.

Having accounted for the mass range of the exotic Higgs bosons, 
which is quite relevant for the LHC,
we now write down\cite{Davidson:2005cw, Altunkaynak:2015twa} 
the Yukawa couplings:\footnote{
 Similar formulas were given in Ref.~\refcite{Mahmoudi:2009zx},
 but this reference assumed the {\boldmath $\rho$}$^f$s are diagonal.
}
\begin{align}
    - \frac{1}{\sqrt{2}} \sum_{f = u, d, \ell} 
 \bar f_{i} &\Big[\big(-\lambda^f_i \delta_{ij} s_\gamma + \rho^f_{ij} c_\gamma\big) h
  + \big(\lambda^f_i \delta_{ij} c_\gamma + \rho^f_{ij} s_\gamma\big)H
  - i\,{\rm sgn}(Q_f) \rho^f_{ij} A\Big]  R\, f_{j} \notag\\
  - \bar{u}_i&\left[(V\rho^d)_{ij} R-(\rho^{u\dagger}V)_{ij} L\right]d_j H^+ 
 - \bar{\nu}_i\rho^\ell_{ij} R \, \ell_j H^+
 +{h.c.},
\label{eq:Yuk}
\end{align}
where the generation indices $i$, $j$ are summed over, 
$L, R = (1\mp\gamma_5)/2$ are projection operators, 
and $V$ is the Cabibbo-Kobayashi-Maskawa matrix,
with the corresponding matrix in lepton sector taken as unity.\footnote{
 This holds true as the active neutrinos are rather degenerate
 at vanishingly small masses.
}
The {\boldmath $\lambda$}$^f$ matrices have been diagonalized as usual 
with elements $\lambda_i^f = \sqrt2 m_i^f/v$,
but the extra Yukawa matrices, {\boldmath $\rho$}$^f$,
cannot be diagonalized simultaneously in principle.
As has been said, however, the mass-mixing structure that Nature has revealed to us, 
together with the decoupling of $h$ from {\boldmath $\rho$}$^f$ matrices 
in the alignment limit ($c_\gamma \to 0$), seem sufficient to shield 
FCNH couplings from our view,
which could be the ``flavor'' design of Nature.

It should be stressed that, not only the {\boldmath $\rho$}$^f$ matrices
cannot be simultaneously diagonalized with the {\boldmath $\lambda$}$^f$ matrices,
they are in principle complex, with\cite{Hou:1991un} $\rho_{tc}$ 
and\cite{Chen:2013qta} $\rho_{tt}$ expected to be the largest elements.
In fact, $\rho_{tt}$ being ${\cal O}(1)$ and complex is the most plausible,
as it is the companion to the diagonal $\lambda_t \cong 1$, {\it which has been recently 
affirmed\cite{Sirunyan:2018hoz, Aaboud:2018urx} by experiment}. 
Exploiting this, it was shown that electroweak baryogenesis (EWBG) 
can be achieved,\cite{Fuyuto:2017ewj} with the $CP$ violating (CPV) strength,
\begin{align}
 \lambda_t{\rm Im}\, \rho_{tt} = {\cal O}(1),
 \label{eq:CPVBAU}
\end{align}
as the driver. Interestingly, the ``naturalness'' condition of ${\cal O}(1)$ 
dimensionless parameters in the Higgs potential is precisely what is needed to 
allow\cite{Kanemura:2004ch} a first order phase transition, a prerequisite for EWBG.
On the other hand, with the recent order of magnitude improvement of 
the limit on the electron electric dipole moment (eEDM) by 
ACME18,\cite{Andreev:2018ayy} the EWBG study has been 
updated\cite{Fuyuto:2019svr} by adding the previously ignored $\rho_{ee}$ 
as a complex parameter. It is found that, so long that 
\begin{align}
  \rho_{ee}/\rho_{tt} \propto \lambda_e/\lambda_t,
 \label{eq:rhoee-rhott}
\end{align}
holds, which rhymes with the mass-mixing hierarchy, 
eEDM could be suppressed by two orders of magnitude or even more,
and there can be an ACME (or other competing experiment) discovery 
in the not so distant future.

It can now be said that g2HDM touches ``the Heavens and the Earth''!
%
%
We now turn to see how extra Yukawa couplings can be explored at the LHC.
We retrace in Sec.~3 the saga of the ongoing $t \to ch$ search,
then on to our main theme of direct production processes 
via TCNH couplings in Sec.~4.

\section{Prelude: $t \to ch$ search}

We have already given a brief survey of theory work on $t \to ch$
in the Introduction. But with $h(125)$ found lighter than top, 
one enters a different era.
The search\footnote{
 We will keep to $tch$ couplings in our main text, and
 comment on search for $tuh$ couplings, where constraint is slightly weaker,
 in Sec.~5.
}
for $t \to ch$ decay commenced immediately with the $h(125)$ discovery.
Even if NFC is operative in Nature, it is a no-lose situation for pursuing
the experimental search. It is in fact an obligation.

It should be clear  that ${\cal B}(t \to ch)$ cannot be overly large, 
otherwise even Tevatron $t\bar t$ studies might have uncovered it.
At the LHC, a multi-lepton analysis\cite{Craig:2012vj} using CMS 7 TeV 
results\cite{Chatrchyan:2012mea} gave a bound at 2.7\% already in 2012.
It was cautioned,\cite{Chen:2013qta} however, 
that one needs to take possible modifications of $h$ properties into account.
It was then\cite{Chen:2013qta} stressed that the $h(125)$ discovery events 
in $ZZ^*$ final states themselves {\it could} contain information 
on ${\cal B}(t \to ch)$ because of its sizable branching fraction
and exceptional cleanliness. 
For example, one should look for accompanying jets enriched with $b$s,
if the events actually cascaded down from $t\bar t$ production. 
The conclusion was that 2011-2012 data should be able to reach the 1\% level.
This carried with it some sense of excitement, as discovery was possible. 
Indeed, the experimental measurements were already under way. 

By 2013 summer conferences, ATLAS reported\cite{ATLAS:2013nia} 
a result on $t \to ch$ search via $h \to \gamma\gamma$ with 20 fb$^{-1}$ data at 8 TeV.
The published limit\cite{Aad:2014dya} of $0.79\%$ in 2014 at 95\% C.L.
indeed broke the 1\% floor. 
Using same amount of data and combining $\gamma\gamma$ and
multi-lepton results, CMS found\cite{Khachatryan:2014jya} a better limit at $0.56\%$.

We refrain from accounting for further developments, as it is
well documented in PDG,\cite{PDG}
but comment that $t \to ch$ search via the dominant $h \to b\bar b$ decay 
has yet to deliver its full\cite{Kao:2011aa} promise,
and should be explored further.
The current best limit on $t \to ch$ search at 95\% C.L. is 
from ATLAS,\cite{Aaboud:2018oqm}
\begin{align}
  {\cal B}(t \to ch) < 1.1  \times 10^{-3},\quad\quad ({\rm ATLAS\;36\;fb}^{-1},\;2019)
 \label{eq:Btch}
\end{align}
which combines the $h \to bb$, $\tau\tau$ modes with earlier
$h \to WW,\, ZZ$, and $\gamma\gamma$ Run~2 results.
The two photon mode gave a limit\cite{Aaboud:2017mfd} at $2.2 \times 10^{-3}$, 
but remarkably, the limit from the $\tau\tau$ mode\cite{Aaboud:2018oqm} 
at $1.9 \times 10^{-3}$ is better. 
As this is only a fraction of  Run~2 data, the bound would 
continue to improve in the near future,
while CMS should certainly be watched also.

Although experiments continue to set constraints on the
effective $\lambda_{tch}$ Yukawa coupling, 
Ref.~\refcite{Chen:2013qta} pointed out that the coupling in fact should be
\begin{align}
\lambda_{tch} = \rho_{tc} c_\gamma,
 \label{eq:tch-Yuk}
\end{align}
in g2HDM, where $\rho_{tc}$ is a TCNH coupling from 
the extra Yukawa matrix {\boldmath $\rho$}$^u$,
defined in Eq,~(\ref{eq:Yuk}). Since we know that $c_\gamma \equiv \cos\gamma$
 (denoted usually as $\cos(\beta-\alpha)$ in 2HDMs with $Z_2$) 
has to be small, the nonobservation so far can be accounted for, 
without requiring $\rho_{tc}$ to be small, even though 
discovery can still happen at any time.\footnote{
 In contrast, one does not quite expect $h \to \tau\mu$ to be seen,
 as the coupling $\rho_{\tau\mu}c_\gamma$ is suppressed by
 both $\rho_{\tau\mu} \lesssim \lambda_\tau \ll \rho_{tc}$
 (most likely) as well as the alignment parameter $c_\gamma$.
 There was once a CMS hint from Run 1 data, but it subsequently
 disappeared by adding data.\cite{PDG}
}
Note that we have dropped $\rho_{ct}$ in Eq.~(\ref{eq:tch-Yuk}), as 
it has to be rather small,\cite{Altunkaynak:2015twa} 
because its effect in $B_d$ and $B_s$ mixing is CKM enhanced.

\section{Probing Extra Top Yukawa Couplings in Production Processes}

From the emergent ``alignment'' phenomenon,\cite{Khachatryan:2016vau} 
we learned that Nature has further designs\cite{Hou:2017hiw} for suppressing
FCNH effects at low energy: small $c_\gamma$, or alignment in the Higgs sector.
This is reflected in Eq.~(\ref{eq:tch-Yuk}), which can be read off from Eq.~(\ref{eq:Yuk}).
It applies also to $h \to \tau\mu$, where there was once a hint\cite{PDG} from 
CMS Run~1 data, but not supported by Run~2 data (see footnote j below).
To probe the TCNH coupling $\rho_{tc}$ without $c_\gamma$ suppression, 
one has to access the direct production of the $H$, $A$ and $H^+$ bosons,
where the mass range in Eq.~(\ref{eq:mrange}) that follows from naturalness
(${\cal O}(1)$ parameters) seems tailor-made for the LHC.
We turn to such processes, namely\cite{Kohda:2017fkn} 
$cg \to tH, tA$ and\cite{Ghosh:2019exx} $cg \to bH^+$,  in this main section.

The existence of the extra diagonal Yukawa coupling $\rho_{tt}$ means
$H$, $A$ can be produced by gluon-gluon fusion. 
We relegate resonance production, namely 
$gg \to H,\,A \to t\bar t,\, t\bar c$ and $\tau\mu$ to Sec.~5,
as they face various and different challenges.

\subsection{Top-associated $H/A$ Production: $cg \to tH/tA \to tt\bar c$, $tt\bar t$}

\begin{figure}[b]
\center
\includegraphics[width=.5\textwidth]{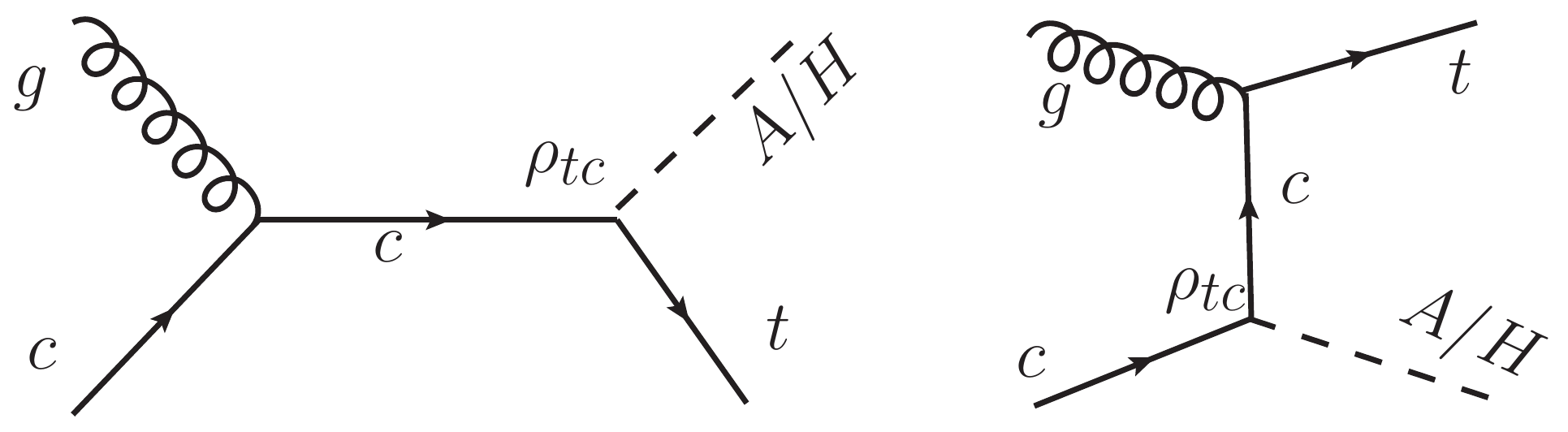}
\caption{
 The $cg \to tS$ ($S = H, A$) process.
}
\label{tAtH}
\end{figure}

We first consider $cg \to tH$, $tA$ production 
via the TCNH $\rho_{tc}$ coupling (see Fig.~\ref{tAtH}), 
with subsequent decay of $H$,\,$A$ to $t\bar c$ and $t\bar t$ final states, 
which depends on $H$,\,$A$ masses 
and the strength of the extra diagonal $\rho_{tt}$ coupling.

\begin{figure}[t]
\center
\includegraphics[width=.45\textwidth]{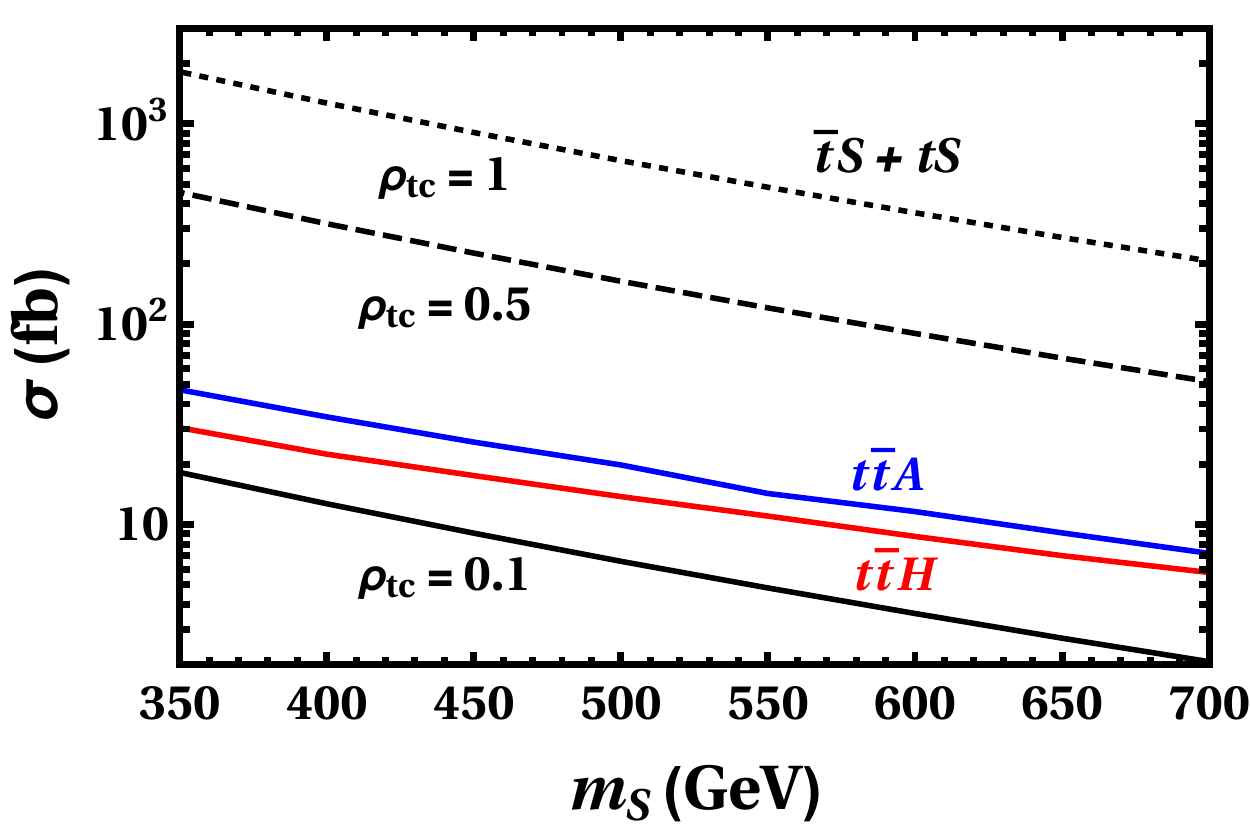}
\caption{
 Cross sections $\sigma(pp \to tS)$ and $\sigma(pp \to t\bar tS)$ for
 $\rho_{tt} = 1$ and $\rho_{tc} = 0.1$ (solid), 0.5 (dashed) and 1 (dots)
 at $\sqrt s = 14$ TeV.\cite{Kohda:2017fkn}
}
\label{partcross}
\end{figure}

\begin{figure}[t]
\center
\includegraphics[width=.4 \textwidth]{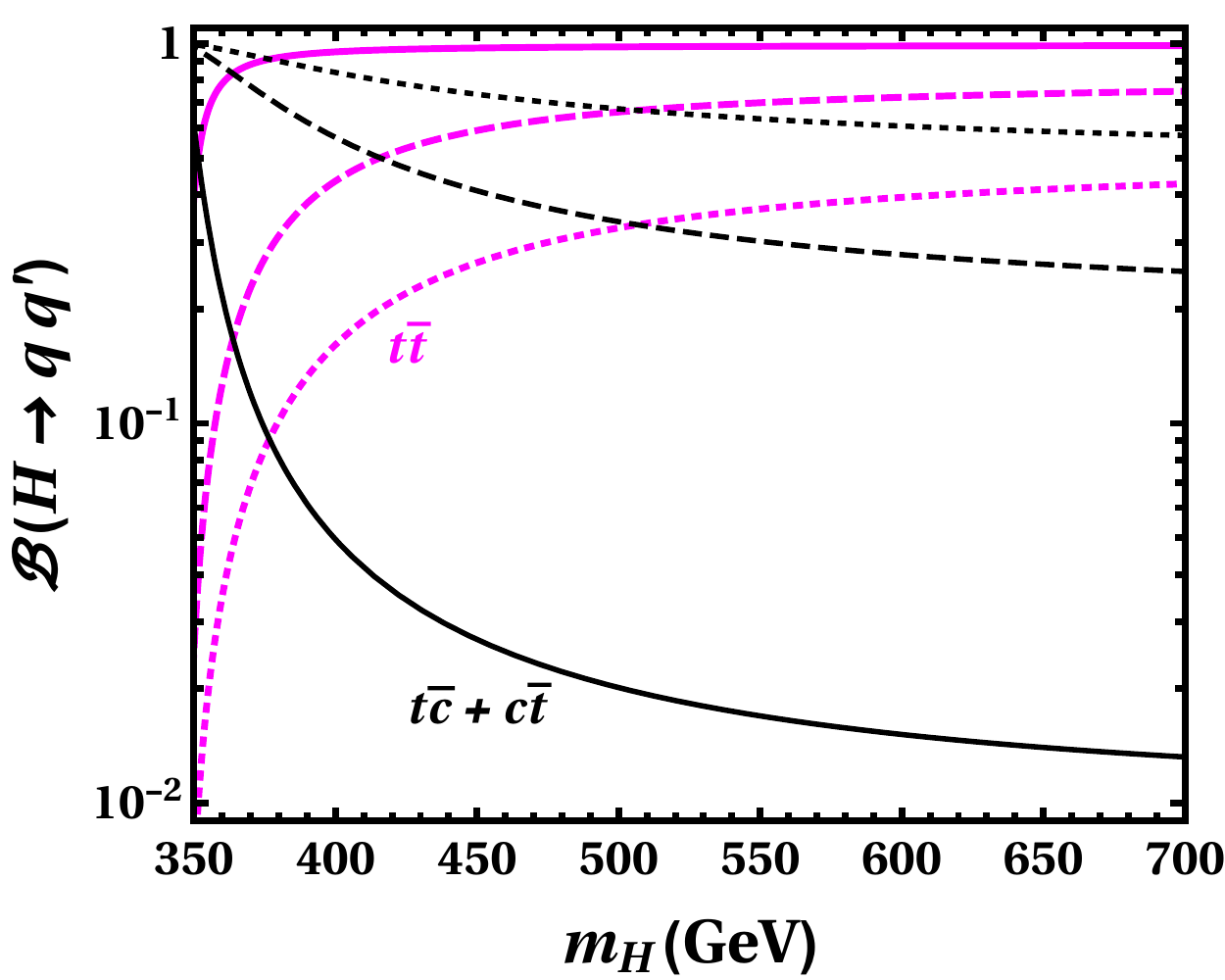}
\includegraphics[width=.4 \textwidth]{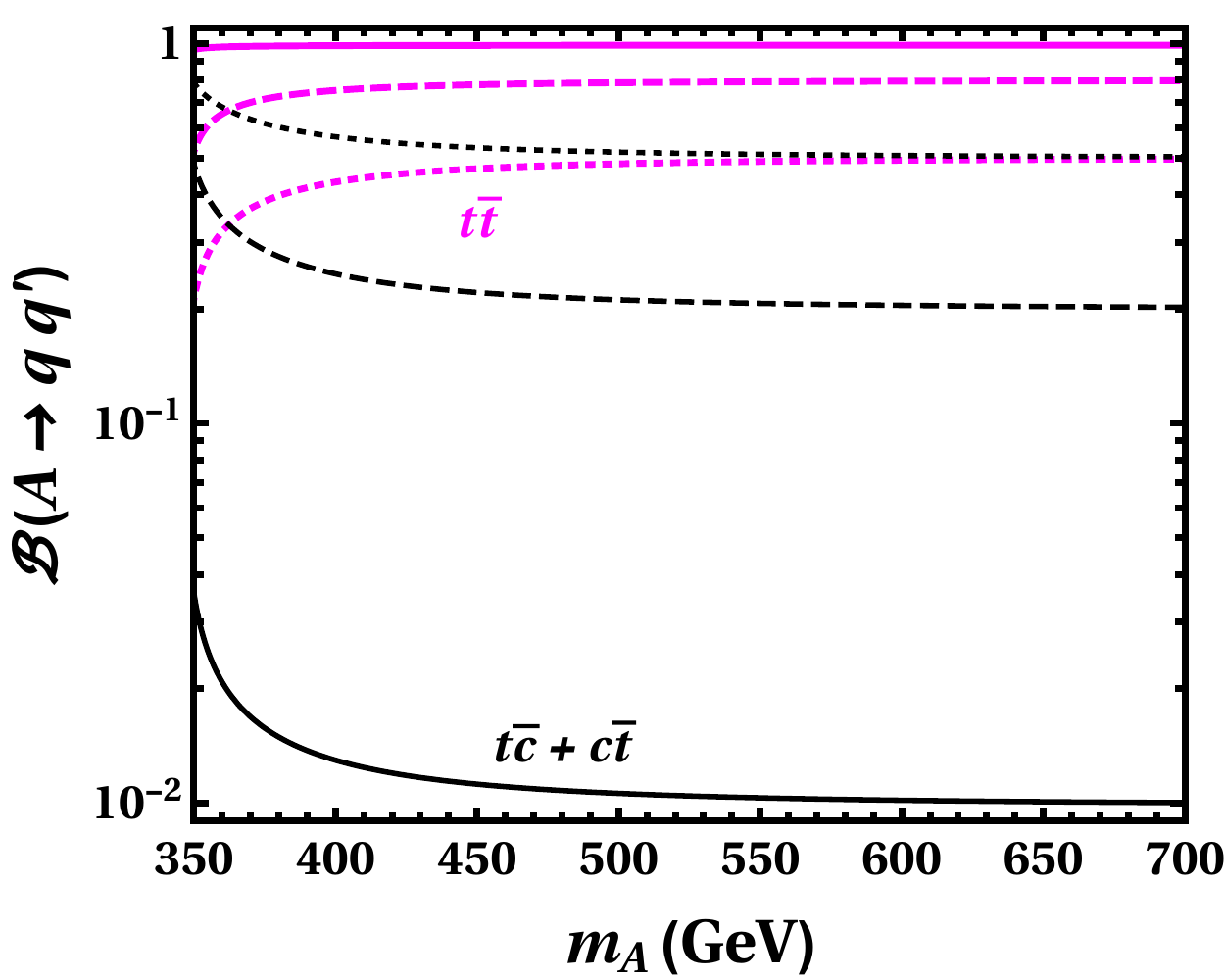}
\caption{
 ${\cal B}(H/A \to t\bar t, t\bar c)$ for $\rho_{tt} = 1$ and $\rho_{tc}
 = 0.1$ (solid), 0.5 (dashed) and 1 (dots).\cite{Kohda:2017fkn}
}
\label{width0}
\end{figure}

To compute the decay rates and parton cross sections, we simplify and 
take the alignment limit of $c_\gamma = 0$. 
The extra Yukawa couplings for $u$-type quarks are,
\begin{align}
\frac{\rho_{ij}}{\sqrt2}\, \bar u_{iL}(H + iA)u_{jR} + {\rm h.c.},
 \quad\quad\ (c_\gamma = 0)
 \label{eq:rho_ij-u}
\end{align}
where $\rho_{ij}$ should share the ``flavor organization'' attributes
of SM, i.e. trickling down off-diagonal elements.
For sake of discussion, we keep only $\rho_{tc}$ and $\rho_{tt}$ finite
($\rho_{ct}$ is constrained small\cite{Altunkaynak:2015twa} by $B$ physics) 
and set all other $\rho_{ij}$s to zero, including those of down and lepton sectors, 
for the remainder of this section.
For $S = H,\, A$, we display\cite{Kohda:2017fkn} $\sigma(cg \to tS)$ 
at parton level in Fig.~\ref{partcross} for $m_{S} \in (350,\, 700)$ GeV.
The $H, A$ decay branching fractions are given in Fig.~\ref{width0}.

%
We will see that\cite{Kohda:2017fkn} $cg \to tS \to tt\bar c$, 
or same-sign top pair (SS$2t$) production,\footnote{
 Same-sign top was advocated\cite{Berger:2011ua} to appear with 
 even 1~fb$^{-1}$ LHC data due to some Tevatron ``anomaly'',
 but this was quickly ruled out.\cite{Aad:2012bb, Chatrchyan:2012sa}
 Same-sign top production has also been advocated recently
 with flavor-changing $Z'$,\cite{Ebadi:2018ueq,Cho:2019stk} 
 as well as neutral scalar $\phi$ exchange.\cite{Ebadi:2018ueq}}
is already quite promising with 300\,fb$^{-1}$ 
if $S$ is in the lower range of  Eq.~(\ref{eq:mrange}). 
For the higher mass range,
i.e. $S$ considerably above the $2m_t \gtrsim 350$~GeV threshold,
the $tt\bar t$ ($3t$) signature holds more promise 
in cutting down background at the High Luminosity LHC (HL-LHC).
Note that the $3t$ cross section in SM is at ${\cal O}$(fb),\cite{Barger:2010uw}
while g2HDM could enhance this by a factor of several hundred
 (cf. Fig.~\ref{partcross}).
Although both CMS\cite{Sirunyan:2019wxt} and ATLAS\cite{Aad:2020klt} 
have zoomed in on $tt\bar t\bar t$ (i.e. $4t$) search\footnote{
 This is in part because SM $4t$ cross section\cite{Barger:2010uw,Frederix:2017wme}
 at $\sim 12$ fb is an order of magnitude larger than $3t$ in SM,
 hence can be measured sooner if SM holds true.
} 
with full Run~2 data, neither experiments have initiated studies targeting $3t$ so far.
We give in Fig.~\ref{partcross} also the $t\bar tS$ cross section,
where $t\bar tA$ is larger than $t\bar tH$.
While almost two orders below $tS$ for $\rho_{tc} \simeq 1$,
it can feed our signatures,
and would dominate over $tS$ for low $\rho_{tc}$.\footnote{
{We do not include~\cite{Gori:2017tvg} $t \bar c S$ in Fig.~\ref{partcross}.
However, it is included in our collider signal for the inclusive same-sign top 
(marked as SS$2t$) production discussed in Sec.~4.1.2.}
}

\subsubsection{Collider Constraints on $\rho_{tc}$}

A small $c_\gamma$ makes the $t \to ch$ constraint on $\rho_{tc}$ rather mute. 
But since the original proposal\cite{Kohda:2017fkn} for  $cg \to tH/tA \to tt\bar c$ search,
both CMS\cite{Sirunyan:2019wxt} and ATLAS\cite{Aad:2020klt} 
have now searched for $4t$ production with full Run~2 data set of 137~fb$^{-1}$ at 13 TeV.
Some of the Control Regions (CRs) of these $4t$ studies
can probe into the SS$2t$ signature, which can be used to constrain $\rho_{tc}$,
independent of $c_\gamma$.
The results of this and the following subsection therefore follow
the more recent analysis of Ref.~\refcite{Hou:2020ciy}, updating
from Ref.~\refcite{Kohda:2017fkn} for the $\rho_{tc}$ study.

Let us start with the CMS $4t$ search.\cite{Sirunyan:2019wxt}
With the baseline selection criterion of at least two same-sign leptons,
we find the most stringent constraint on $\rho_{tc}$ 
arises from CRW~\cite{Sirunyan:2019wxt}, 
the CR of $t\bar t W$. 
The CRW of the CMS $4t$ search~\cite{Sirunyan:2019wxt} is defined as 
containing two same-sign leptons plus two to five jets with two $b$-tagged.
The selection cuts are as follows.
For leading\,(subleading) lepton transverse momentum,
 $p^{\ell_1\,(\ell_2)}_T$\,$>$\,$25$\,(20)\;GeV. 
For pseudorapidity, $\eta^e < 2.5$ while $\eta^{\mu,\, j} < 2.4$.
The $p_T$ of ($b$-)jets should satisfy 
one of the following: 
(i) $p_T$\,$>$\,$40$\;GeV for both $b$-jets;
(ii)~$p_T$\,$>$\,$20$\;GeV for one $b$-jet,
 $20$\,$<$\,$p_T$\,$<$\,$40$\;GeV for the second $b$-jet, and
 $p_T$\,$>$\,$40$\;GeV for the third jet;
(iii) $20$\,$<$\,$p_T$\,$<$\,$40$\;GeV for both $b$-jets,
 with two extra jets each satisfying $p_T$\,$>$\,$40$\;GeV.
Defined as the scalar sum of $p_T$ of all jets, one requires
$H_T$\,$>$\,$300$\;GeV, while $p_T^{\rm miss}$\,$>$\,$ 50$\;GeV.
To reduce the Drell-Yan background with a charge-misidentified ($Q$-flip) electron,
events with same-sign electron pairs with $m_{ee}$\,$<$\,$12$\;GeV are rejected.
Using these selection cuts, CMS reports 338 observed events in CRW,
with $335 \pm 18$ events (SM backgrounds plus $4t$) expected~\cite{Sirunyan:2019wxt}.

To estimate our limits, we use MadGraph5\_aMC@NLO\cite{Alwall:2014hca}
 (denoted as MadGraph5\_aMC) to generate signal events {for $\sqrt{s}=13$ TeV} at leading order (LO) 
with default parton distribution function (PDF) set NN23LO1~\cite{Ball:2013hta}, 
{interface} with PYTHIA~6.4~\cite{Sjostrand:2006za}
 for showering and hadronization, 
and MLM matching\cite{Mangano:2006rw, Alwall:2007fs} prescription
 for matrix element (ME) and parton shower merging.
The event samples are then fed into DELPHES~3.4.2~\cite{deFavereau:2013fsa}
{for fast detector simulation. 
We utilize the default $b$-tagging efficiency and light-jet rejection 
of DELPHES CMS-based detector card.
The jets are reconstructed via anti-$k_T$ algorithm with radius parameter $R=0.6$, 
whereas}
the effective model is implemented in FeynRules~\cite{Alloul:2013bka}.

\begin{figure*}[t]
\center
\includegraphics[width=.48 \textwidth]{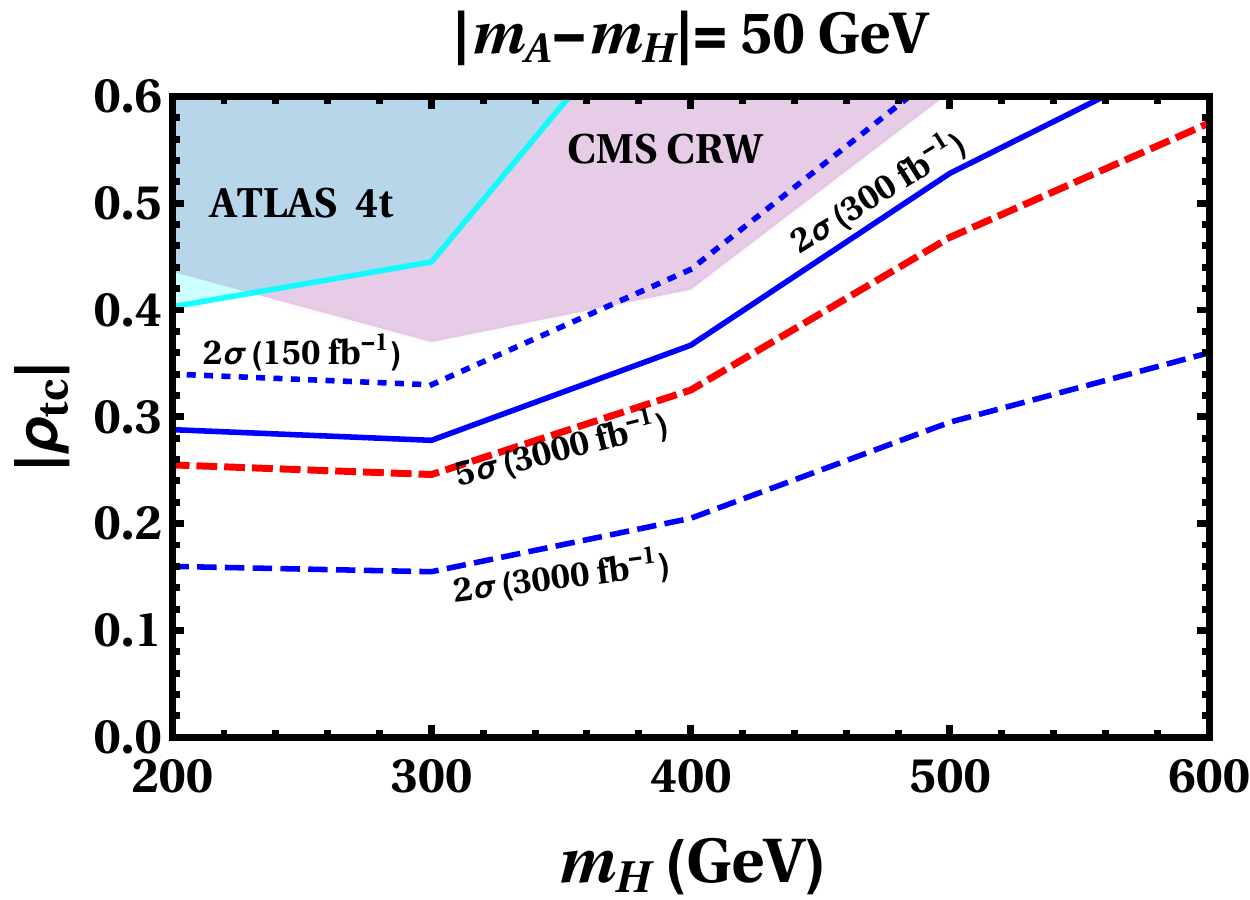}
\includegraphics[width=.48 \textwidth]{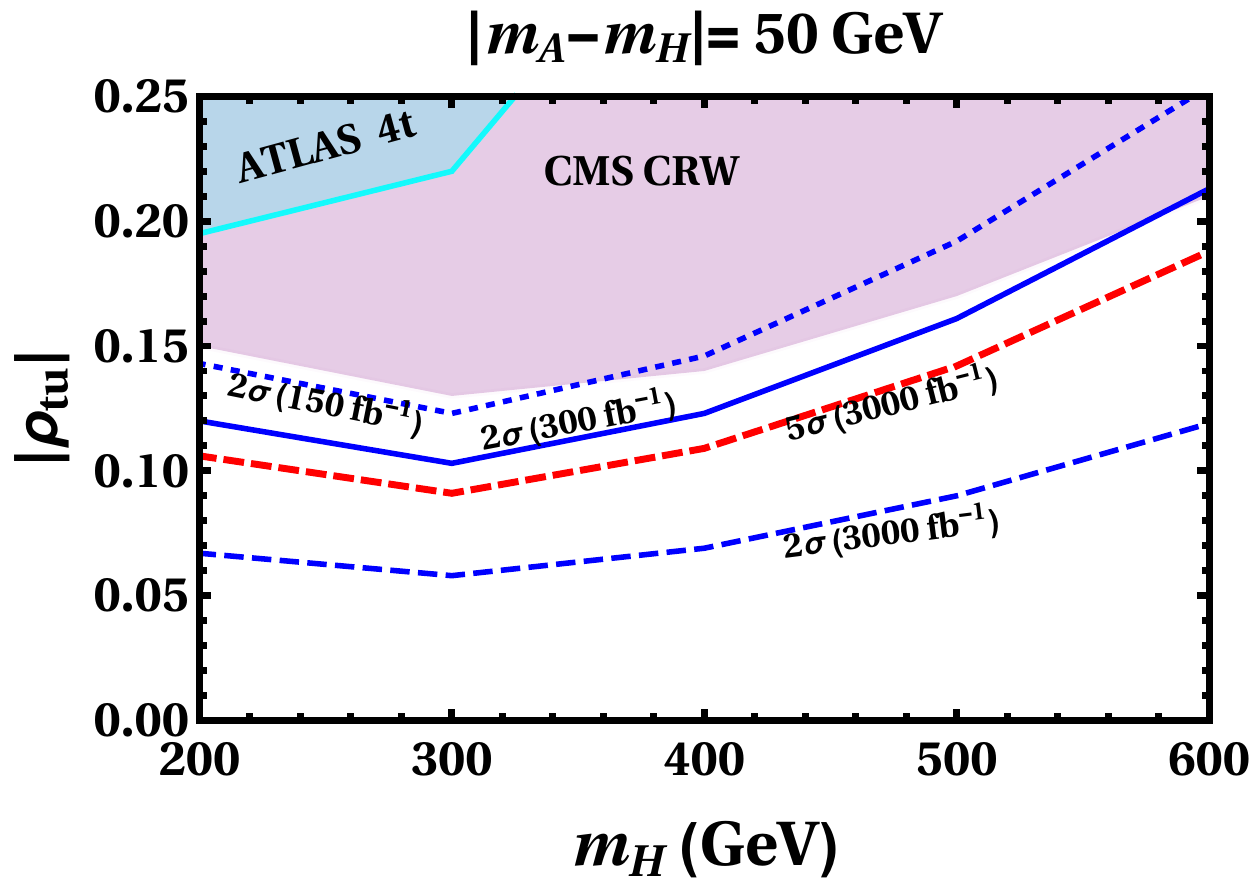}
\caption{
[left] 
Constraint from $4t$ search in $m_H$--$\rho_{tc}$ plane
 (with $m_A - m_H = 50$~GeV)
 by CRW of CMS\cite{Sirunyan:2019wxt} (purple)
 and CRttW2$\ell$ of ATLAS\cite{Aad:2020klt} (cyan),
setting all other $\rho_{ij} = 0$. 
Plotted further are the exclusion limits (blue) and discovery reaches (red)
 for $\rho_{tc}$ from same-sign top signature
 for various integrated luminosities at the 14 TeV LHC.
[right] 
The same for $\rho_{tu}$.\cite{Hou:2020ciy}
}
 \label{exclu}
\end{figure*}

The 
$pp\to tH/tA \to t t \bar c$ process
with both top quarks decaying semileptonically 
(nonresonant $cg \to t t \bar c$,
$t$-channel $H/A$ exchange $cc\to tt$ and $gg\to t \bar c A/H$ processes are included) 
contributes to CRW of the CMS $4t$ search. 
Setting all other $\rho_{ij} = 0$, we estimate 
the cross section 
for $\rho_{tc}=1$ then scale by $\rho_{tc}^2$, 
assuming narrow $H/A$ widths with {$\mathcal{B}(H/A\to t \bar c +\bar t c) = 100\%$}. 
We then demand the sum of expected SM 
and $\rho_{tc}$-induced events agree with observed, 
where the $2\sigma$ excluded region from CRW is displayed 
in Fig.~\ref{exclu}[left] in purple, assuming Gaussian behavior for simplicity.
To avoid cancellation\cite{Kohda:2017fkn} between 
$H$ and $A$ mediated production processes,
we have assumed a splitting of $m_A - m_H = 50$~GeV.
This almost doubles the production rate, while if one of the exotic
boson is much heavier than the other, the cross section drops by roughly one half.
%
%
One can see the power of the $4t$ search, constraining $\rho_{tc}$
especially for $m_S$ not far above the $t\bar t$ threshold of 350~GeV,
but tapers off rapidly for heavier Higgs bosons {due to rapid fall in the parton
luminosity}.

ATLAS has also searched for $4t$ production~\cite{Aad:2020klt}
with 139~fb$^{-1}$, but categorizing into different CRs and Signal Regions (SRs). 
Again the CR for $t\bar t W$, called $\rm{CRttW2\ell}$, is the most relevant. 
It is defined as
 at least two same-sign leptons ($e\mu$ or $\mu\mu$),
 plus at least four jets with at least two $b$-tagged, where 
$p_T > 28$\;GeV with $\eta^\mu < 2.5$ and $\eta^e < 1.5$ 
for the same-sign leptons. 
All jets should satisfy $p_T > 25$\;GeV and $\eta < 2.5$.
For two $b$-jets, or three or more $b$-jets but with no more than 5 jets,
 the scalar $p_T$ sum over all jets and same-sign leptons
 (different from CMS), $H_T < 500$\;GeV. 
Unlike CRW for CMS, 
ATLAS does not give the observed number of events in $\rm{CRttW2\ell}$,
but provides a figure of comparison between data and prediction 
in the variable $\sum p_T^\ell$ (see Ref.~\refcite{Aad:2020klt} for definition).
We follow Ref.~\refcite{Hou:2018npi} and digitize the figure to extract 
the number of {expected and observed events in the $\rm{CRttW2\ell}$ 
from the $\sum p_T^\ell$ distribution}.
We find $378\pm 10$ and $380$, respectively,
where we simply add the errors in quadrature for the expected events 
from each $\sum p_T^\ell$ bin.

To extract the constraint, we follow the event selection as described 
and use now the ATLAS-based detector card of DELPHES.
Assuming that the $\rho_{tc}$-induced events plus SM stay 
within $2\sigma$ of expected, 
the exclusion limits from ATLAS 
are the cyan shaded region in Fig.~\ref{exclu}[left],
which is weaker due to different selection cuts.
From CMS $4t$ search, we find {$\rho_{tc} \lesssim 0.37$--$0.44$}
is still allowed for {$200\,\mbox{GeV}\lesssim m_H \lesssim 400$\;GeV}, 
while larger values open up quickly for $m_H > 400$\;GeV.
We again illustrate for $m_H - m_A = 50$\;GeV, 
as there is strong cancellation\cite{Kohda:2017fkn} between
$cg \to tH \to tt\bar c$ and $cg \to tA \to tt\bar c$ amplitudes
for $H,\,A$ {that are nearly degenerate in mass and width}.
Note also that SUSY and other exotic searches nowadays use rather strong cuts 
and do not give relevant constraints for our relatively low mass range.

Fig.~\ref{exclu}[right] is done in the same way as the left-hand plot,
but with $\rho_{tc}$ replaced\cite{Hou:2020ciy} by $\rho_{tu}$,
which we would discuss only in Sec.~5.

\subsubsection{Same-sign Top:  $cg \to tH/tA \to tt\bar c$}

Although existing\ $4t$ searches with full LHC Run~2 data can constrain 
$\rho_{tc}$, they are not optimized for $cg \to tH/tA \to tt\bar c$ search.
We advocate\cite{Kohda:2017fkn} a dedicated 
study of a stand-alone $\rho_{tc}$ coupling with all other $\rho_{ij} = 0$.

The $pp \to tH/tA + X \to tt\bar c + X$ signature is defined as
same-sign dilepton ($ee$, $e\mu$, $\mu\mu$) plus at least three jets
with at least two $b$-tagged and one non-$b$-tagged,
and 
$E_T^{\rm miss}$. 
%
The SM backgrounds are $t\bar t Z$, $t\bar t W$, $4t$, $t\bar t h$, 
with $tZ +$ jets subdominant ($3t + j$ and $3t + W$ turn out negligible).
The $t\bar t$ and $Z/\gamma^* +$ jets processes, with sizable cross sections, 
would contribute if {one lepton charge is misidentified ($Q$-flip)}, with probability 
taken as\cite{ATLAS:2016kjm,Alvarez:2016nrz,Aaboud:2018xpj}
$2.2\times 10^{-4}$ in our analysis.

As before, we generate signal and background events at LO 
via MadGraph5\_aMC but for $\sqrt{s}=14$\;TeV and follow the same 
showering, hadronization, ME matching and parton shower merging,
but adopt the default ATLAS-based detector card of DELPHES.
The LO  $t\bar{t} W^-$\,($t\bar{t} W^+$), $t\bar t Z$, 
$4t$, $t\bar t h$ and $tZ+$\,jets  cross sections are normalized to 
next-to-leading order (NLO) $K$ factors 
1.35 (1.27),~\cite{Campbell:2012dh} 
1.56,~\cite{Campbell:2013yla} 
2.04,~\cite{Alwall:2014hca} {1.27,~\cite{twikittbarh} and 
1.44~\cite{Alwall:2014hca}} respectively.
We assume the same $K$ factor for $\bar{t}Z+$\,jets background for simplicity.
The $Q$-flip $t\bar t +$\,jets and $Z/\gamma^*+$\,jets backgrounds are 
corrected to NNLO cross sections by 
$1.27$~\cite{twikittbar} and 
$1.84$~\cite{Hou:2017ozb}, respectively, where
we use FEWZ\,3.1\cite{Li:2012wna} to obtain 
the latter factor for $Z/\gamma^*+$\,jets. 
%

To suppress backgrounds and optimize for $pp\to t A/tH + X \to t t\bar c + X$, 
we take a cut based analysis that differs from CRW of CMS $4t$ search.
For leading (subleading) lepton, $p^{\ell_1\,(\ell_2)}_T > 25$ (20)\;GeV,
while $\eta^{\ell_1,\,\ell_2} < 2.5$. 
For all three jets, $p_T> 20$\;GeV and $\eta < 2.5$, 
and we demand $E^{\rm miss}_{T} > 30$\;GeV.
The separation $\Delta R$ between (i) a lepton and any jets ($\Delta R_{\ell j}$), 
(ii) the two $b$-jets ($\Delta R_{bb}$), 
and (iii) any two leptons ($\Delta R_{\ell\ell}$)
should all satisfy $\Delta R > 0.4$. 
Finally, with ATLAS definition of $H_T$, i.e. with $p_T$ of 
the two leading leptons included, we demand $H_T > 300$\;GeV.
%
%
The signal cross sections for select $m_H$ values are given 
in the first column of Table~\ref{backg_ssll},
together with the estimated background cross sections,
where the $K$ factors from LO to NLO are shown in brackets. 
%
There are also ``nonprompt'' backgrounds.
The CMS study of same-sign dilepton (SS$2\ell$) signature\cite{Sirunyan:2017uyt},
{with slightly different cuts than ours},
finds nonprompt background at $\sim1.5$ times the $t\bar{t}W$ background.
These backgrounds are not properly modeled in our Monte Carlo simulations,
and we simply add such backgrounds to the overall background 
at 1.5 times $t\bar t W$ after selection cuts.
A realistic analysis by the experiments can handle this better.

\begin{table}[t]
\tbl{{The signal cross sections of the same-sign top SS2$t$
after selection cuts for different $m_H$ (in parentheses) with 
$m_A = m_H \,+\,50$\,GeV for $\rho_{tc}=1$  at 14\,TeV LHC. 
Various backgrounds cross sections after selection cuts are presented in the third column}, 
where numbers in brackets in second column are LO to NLO $K$ factors.}
{
\begin{tabular}{c | c c c c }
 \toprule
    Signal cross section in fb          &        Backgrounds                 &  Cross section (fb)      \\
    ($m_H$ in GeV)                      &                                      &                         \\  
\hline 
     3.83    (200)      &        $t\bar{t}W$\;[1.35\,(1.27)]          & 1.31               \\
     4.12    (300)      &        $t\bar{t}Z$\;[1.56]                  & 1.97                \\
     2.35    (400)      &        $4t$\;[2.04]                         &  \;\,0.092                  \\
     1.14    (500)      &        {$t\bar t h$}\;[1.27]                & \;\,0.058           \\
     0.75    (600)      &        $Q$-flip\;[1.84/1.27]                & \;\,0.024                     \\
               &        $tZ+$jets\;[1.44]                    & \;\,0.007                     \\
 \botrule
\end{tabular}
\label{backg_ssll}
}
\end{table}

To estimate the exclusion limit ($2\sigma$) and discovery potential ($5\sigma$), 
we utilize the test statistics~\cite{Cowan:2010js}
\begin{align}
Z(nx)=\sqrt{-2\ln\frac{L(nx)}{L(nn)}},
 \label{poissn}
\end{align}
where ${L(nx)} =  e^{-x}x^n/n!$ is the likelihood function of Poisson probabilities 
with $n$ the observed number of events, and $x$ is either the number of events predicted 
by the background-only hypothesis $b$, or signal plus background hypothesis $s+b$.
For exclusion ($s+b$) hypothesis, 
we demand ${Z(bs+b)} \geq 2$ for $2\sigma$,
while for discovery ($b$ hypothesis),
we demand ${Z(s+bb)} \geq 5$ for $5\sigma$.
{The signal cross sections for the reference $\rho_{tc}=1$ value 
after selection cuts are given in Table~\ref{backg_ssll} for 
several $m_H$ values with $m_A-m_H = 50$ GeV, assuming $m_H > m_A$.}
{Utilizing the signal} and background cross sections in Table~\ref{backg_ssll}, 
we present in Fig.~\ref{exclu}[left] the exclusion and discovery contours 
{by scaling the signal cross section with $\rho_{tc}^2\,\mathcal{B}(A/H\to t \bar c + \bar t c)$}
in $m_H$--$\rho_{tc}$ plane for different luminosities. 
{We set all $\rho_{ij}=0$ in generating signal, 
which simply translates to $\mathcal{B}(A/H\to t \bar c + \bar t c) = 100\%$.}
We repeat a similar analysis with stand-alone $\rho_{tu}$
(all other $\rho_{ij} = 0$) replacing $\rho_{tc}$,
i.e. for the $pp\to tA/tH + X \to t t\bar u + X$ process 
induced by the $ug \to tH/tA \to tt\bar u$ at parton level.
The exclusion limits and discovery potential are plotted in Fig.~\ref{exclu}[right].
{In Fig.~\ref{exclu}, we have simply interpolated the contours from 
$m_H$ values given in Table~\ref{backg_ssll},
and analogously for $\rho_{tu}$ case.}

We see from Fig.~\ref{exclu}[left] that CRW of CMS $4t$ search\cite{Sirunyan:2019wxt} 
is suitably powerful: with full Run~2 data, our
target analysis can only probe the region of
$\rho_{tc} \sim 0.34$--0.4, while adding full Run~3 data,
the sensitivity extends to a broader region of parameter space,
but so would the continued $4t$ analysis.
The $5\sigma$ discovery region for 3000 fb$^{-1}$ is 
somewhat better than the 300 fb$^{-1}$ exclusion limit,
and one can follow up on any hint from full Run 2+3 data. 
To be watched is $\rho_{tc} \sim 1$, needed for the
backup mechanism for EWBG,~\cite{Fuyuto:2017ewj} for $m_{H,A}$ at the
higher end of the mass range in Fig.~\ref{exclu}[left].
The high significance found
in Ref.~\refcite{Kohda:2017fkn}
suggests it could already be relevant in LHC Run 2.
Same-sign top pair search has the potential to reveal to us whether 
the $\rho_{tc} \sim 1$ mechanism for EWBG is allowed.

As noted, CMS has searched~\cite{Sirunyan:2017uyt} for 
SS$2\ell$ events using $\sim 36$ fb$^{-1}$ data at 13 TeV, 
which can constrain $\rho_{tc}$. 
It is found\cite{Kohda:2017fkn} that the constraint is only $\rho_{tc} \gtrsim 1$
for $S = H, A$ at 350 GeV, and tapers off to higher values,
so does not affect our discussion above. 
The CMS update\cite{Sirunyan:2020ztc} with full Run~2 data
does not change this conclusion, as cuts have moved upward
hence missing our target mass range.

\subsubsection{Triple-Top:  $cg \to tH/A \to tt\bar t$}

We would advocate,\cite{Kohda:2017fkn} however,  that 
``triple-top'' search is more informative with a larger dataset.
%
 {New physics triple-top search has been advocated in
 topcolor-assisted technicolor model,\cite{Han:2012qu}
 usual 2HDM with $Z_2$ symmetry,\cite{Kanemura:2015nza, Patrick:2017ele}
 a literal effective $tqh$ coupling,\cite{Malekhosseini:2018fgp, Khanpour:2019qnw}
 flavor-changing $Z'$,\cite{Cho:2019stk}
 and also using EFT approach with four-fermi operators 
 involving triple-top.\cite{Chen:2014ewl, Cao:2019qrb, Khatibi:2020mvt}
 Our sub-TeV exotic Higgs bosons have dimension-4 couplings (Sec.~2)
 that render an EFT approach artificial.
} 

As $S\to t\bar t$ decay is needed for $cg \to tS$ to contribute,
it is less sensitive just above $t\bar t$ threshold.
With cross sections smaller due to more exquisite selection cuts,
we give results for 3000\,fb$^{-1}$.
%
{Unlike Sec.~4.1.2 for SS$2t$,
the result presented here have not been updated to take
sufficient account of $4t$ constraint with full Run~2 data.}
{In this subsection, we take $\rho_{tt} = 1$.}

We denote our triple-top signature as $3\ell 3b$, which is defined as:
at least three leptons, 
at least three jets with at least three tagged as $b$-jets, 
plus $E^{\rm miss}_T$.
The selection cuts are: 
for the three leading leptons and $b$-jets, 
$p^{\ell}_{T} > 25$~GeV and $p^{b}_T > 20$~GeV, respectively;
$\eta$, $\Delta R$ and $E_T^{\rm miss}$ are the same as for SS2$t$;
scalar sum, $H_T$, of transverse momenta of
{all three leading leptons and $b$-jets} should satisfy $H_T > 320$~GeV.
To reduce $t\bar t Z \, +$\,jets background,
we veto~\cite{CMS:2017uib} the mass range 
$76~\mbox{GeV}< m_{\ell\ell} < 95~\mbox{GeV}$
for same flavor, opposite charged lepton pairs, and if more than one pair is present, 
the veto is applied to the pair mass closest to $m_Z$.

Dominant SM backgrounds are $t\bar t Z\,+$\,jets and $4t$,
with $t\bar t W+$jets, $t\bar t h$ and $tZ+$jets subdominant
 ($3t +j$, $3t + W$ are less than subdominant).
The $t\bar t\, +$\,jets process can contribute if a
jet gets misidentified as a lepton, with probability taken 
as\cite{Alvarez:2016nrz, Aad:2016tuk} $\epsilon_{\rm fake} = 10^{-4}$.
We do not include nonprompt backgrounds
as they are not properly modeled in Monte Carlo simulations.
Unlike the SS2$t$ signature,
the three hard leptons plus high $b$-jet multiplicity, along with
$Z$-pole veto and $H_T$ cut may reduce such contributions significantly.
The $t\bar t Z \,+$\,jets,  $4t$, {$t\bar t W+$jets}, $t\bar th$ and {$tZ+$jets} cross sections
at LO are adjusted by the same factors {as for SS2$t$},
i.e. 1.56, 2.04, 1.35, 1.27 and 1.44 respectively, and likewise $1.84$
for $t\bar t \,+$\,jets with jet faking lepton. 
Conjugate processes are assumed to have the same correction factors {for simplicity}.

\begin{table}[b]
\tbl{
Backgrounds for $3\ell 3b$ process at 14\;TeV,
where LO to NLO $K$-factors (cross sections with $Z$-pole veto) 
are given in the first (second) parentheses.\cite{Kohda:2017fkn}}
{
\begin{tabular}{c c c c c }
 \toprule
                      Backgrounds                 &  Cross section (fb)      \\
\hline
                      $t\bar t Z+$jets\;(1.56)            & 0.0205~~(0.0026)               \\
                      $4t$~(2.04)                         & 0.0232~~(0.0209)                    \\
                      $t\bar{t}W+$jets\,(1.35)                & 0.0017~~(0.0015)               \\
                      $t\bar t h$~(1.27)           & \;{0.0015~~(0.0013) }                      \\
                      $tZj$+jets~(1.44)                      & \hspace{-.35cm}0.0002\quad\ \ \,(--)                \\
                      $t\bar t$+jets (fake)       & 0.0026~~(0.0025)                     \\
 \botrule
\end{tabular}
\label{backg_3l3b}
}
\end{table}

\begin{figure}[t]
\center
\includegraphics[width=.45 \textwidth]{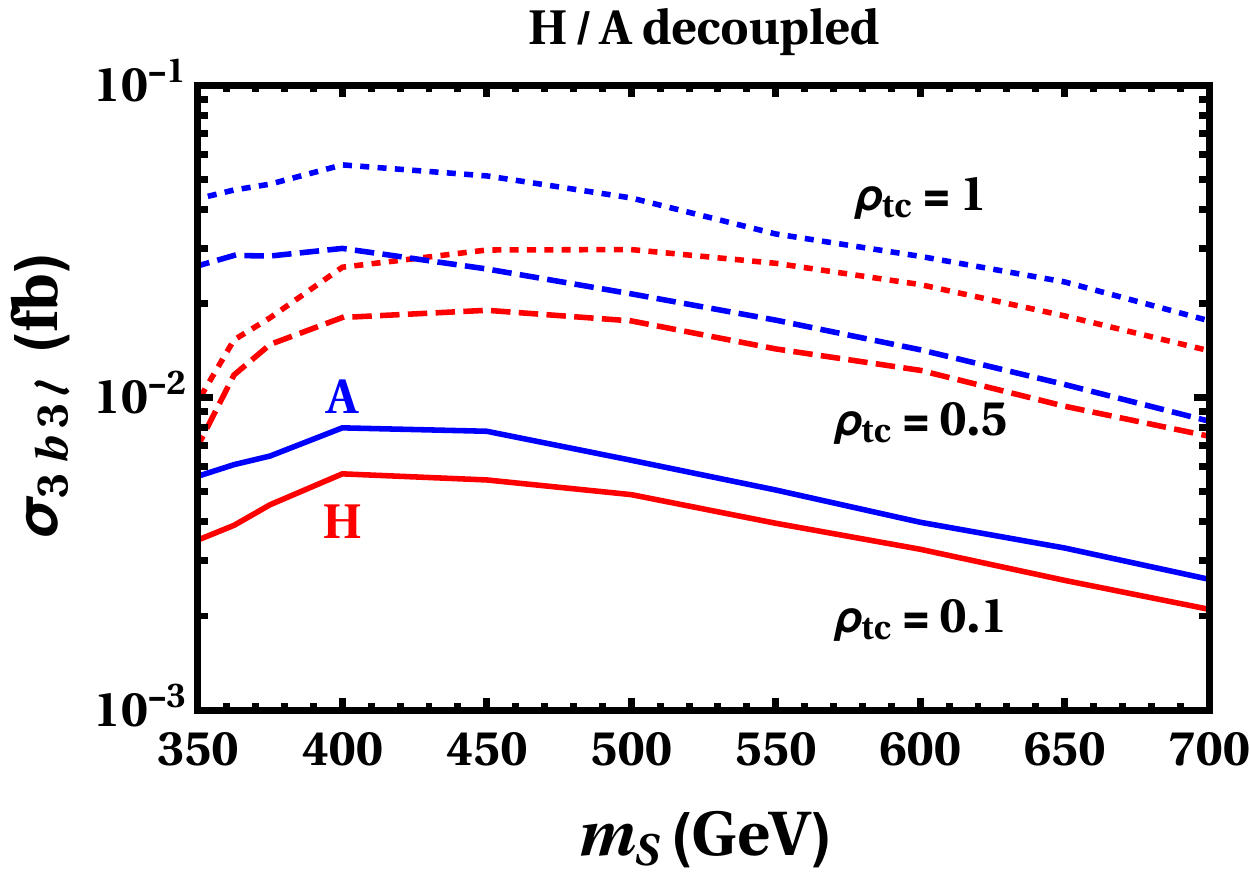}
\includegraphics[width=.44 \textwidth]{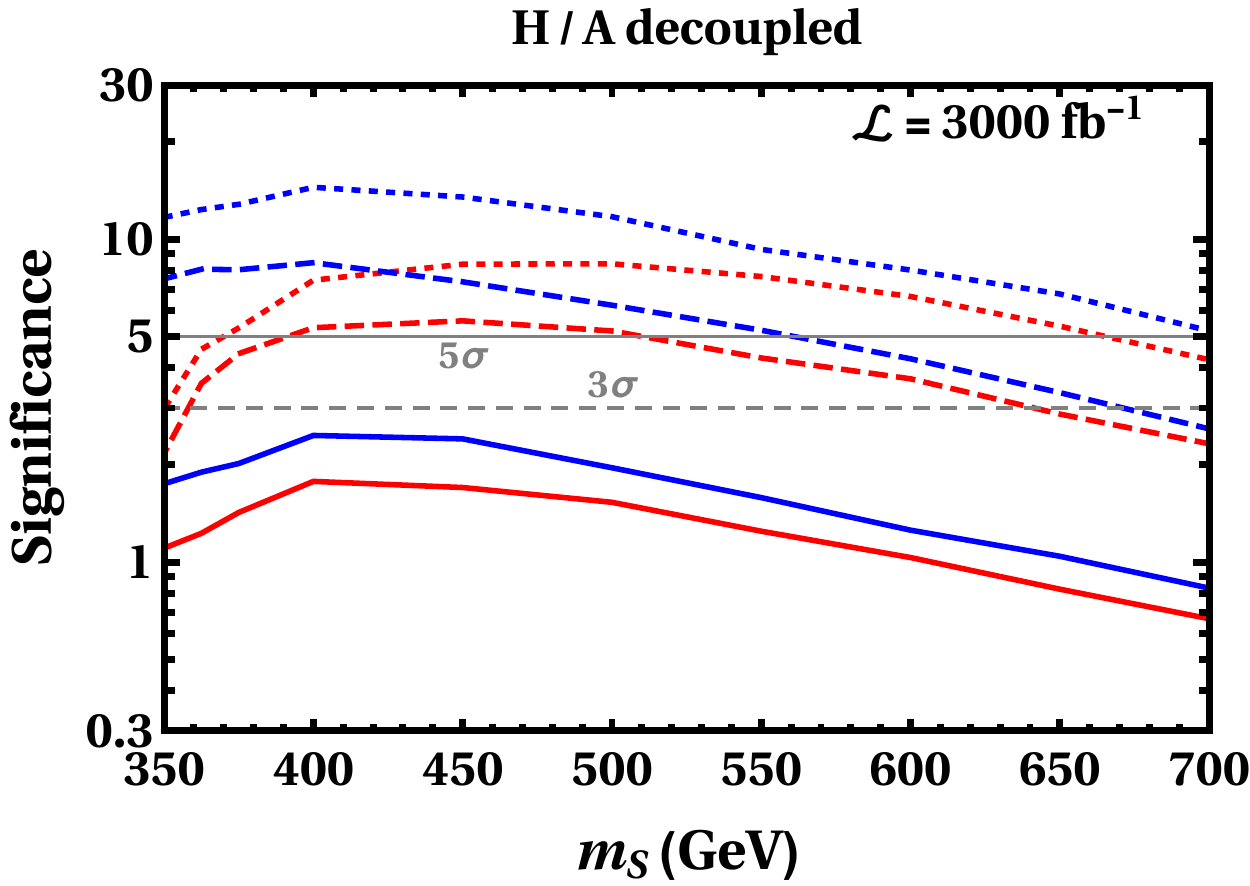}
\caption{
 [left] Cross sections (fb), and [right] significance (at 3000 fb$^{-1}$)
 for $3\ell3b$ final state at $\sqrt s = 14$~TeV for $\rho_{tt} = 1$ and
 several $\rho_{tc}$ values after selection cuts.\cite{Kohda:2017fkn}
}
\label{3b3l}
\end{figure}

With background cross sections given in Table~\ref{backg_3l3b},
the signal cross sections after selection cuts are plotted 
in Fig.~\ref{3b3l}[left] for $\rho_{tt} = 1$ and $\rho_{tc} = 0.1$, 0.5 and 1.
For $\rho_{tc} = 1$ and 0.5, ${\cal B}(A \to t\bar t) > {\cal B}(H \to t\bar t)$ 
(see Fig.~\ref{width0}) makes the cross section for $A$ higher than $H$, 
which also explains the slower turn on for $H$ with $m_H$.
For $\rho_{tc} = 0.1$, the $t\bar tS$ process dominates,
with higher cross section for $A$ (see Fig.~\ref{partcross}).
The {discovery} contours, estimated using Eq.~(\ref{poissn}) for 3000\,fb$^{-1}$,
are plotted in Fig.~\ref{3b3l}[right].
We find $5\sigma$ discovery reach for $\rho_{tc} = 1$ 
covers the full range of Eq.~(\ref{eq:mrange}) and even a bit beyond.
For $\rho_{tc} = 0.5$, $5\sigma$ reach is up to 520 (570)~GeV for $H$\,($A$) 
while $3\sigma$ evidence covers the full range of Eq.~(\ref{eq:mrange}).
{With no interference between $A$ and $H$ induced contributions} for triple-top, 
a small $A$-$H$ mass splitting makes little effect,
hence\cite{Kohda:2017fkn} for degenerate $A$-$H$,
one could make $5\sigma$ discovery even beyond 
the full range of Eq.~(\ref{eq:mrange}).
%

%
%


For lower masses with small splitting, 
the relative strength of triple-top and same-sign top at HL-LHC
could {in principle} allow one to extract information on 
relative strength of $\rho_{tc}$ vs $\rho_{tt}$, 
assuming discovery.\cite{Kohda:2017fkn}
This would throw light on whether EWBG is more 
driven by $\rho_{tt}$ or $\rho_{tc}$.
Although we have elucidated the effect of exotic $H$ and $A$ scalars,
mass reconstruction would not be easy,
and further study would be needed, especially at HL-LHC.
%

We have assumed nearly degenerate heavy scalars, which need not be the case.
Finite splittings could lead to $H \to A Z$, $H^\pm W^\mp$ (or reverse)
decays, which would dilute our signatures but enrich the program.
One charming aspect of extra $\rho_{tt}$ and $\rho_{tc}$ Yukawa couplings 
is their intrinsic complexity, which is why they can drive EWBG~\cite{Fuyuto:2017ewj}.
More studies are needed to probe these CPV phases.
Our proposal is thus only a first step of a large program.

\subsection{Bottom-associated $H^+$ Production: $cg \to bH^+ \to bt\bar b$}

\begin{figure}[t]
\center
\includegraphics[width=0.451\textwidth]{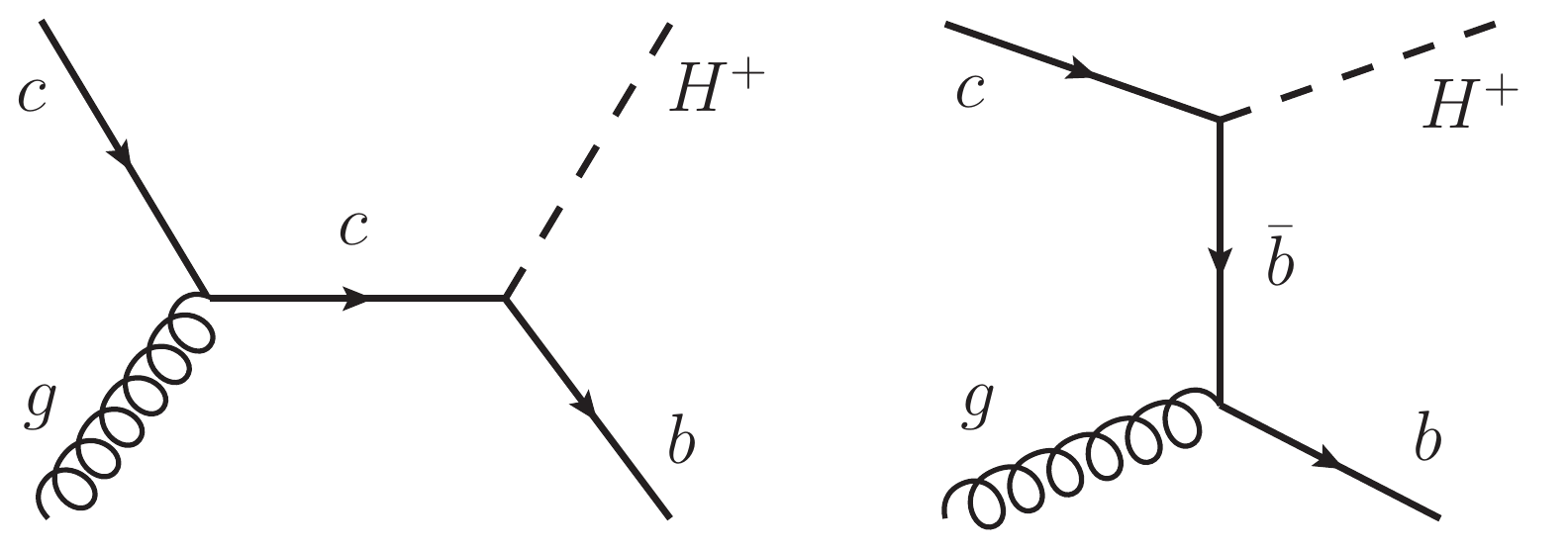}
\caption{Feynman diagrams for $cg\to b H^+$.}
\label{feyndiag}
\end{figure}

A novel process, $cg \to bH^+$ (see Fig.~\ref{feyndiag})
followed by $H^+ \to t\bar b$, was proposed recently,\cite{Ghosh:2019exx} 
that may lead to the discovery of {the exotic} $H^+$ boson in the near future.

The usual 2HDM II motivates the study of the $\bar bg \to \bar tH^+$ process~\cite{Aaboud:2018cwk,Sirunyan:2019arl} 
which goes through the $\bar t bH^+$ coupling.
In contrast, despite being phase space favored, 
the $cg \to bH^+$ process is suppressed 
by the Cabibbo-Kobayashi-Maskawa (CKM) matrix element ratio 
$V_{cb}/V_{tb}^2 \sim 1.6 \times 10^{-3}$.
But in g2HDM with {\it extra} Yukawa couplings,\footnote{
 The $\bar cbH^+$ couplings in 2HDM~III context 
 has been studied in different decay channels and involving various flavor 
assumptions,\cite{He:1998ie, DiazCruz:2009ek, HernandezSanchez:2012eg, Flores-Sanchez:2018dsr, Hernandez-Sanchez:2020vax}
 where our list is only partial.
}
$\bar cbH^+$ and $\bar tbH^+$ couple with strength\cite{Ghosh:2019exx} 
$\rho_{tc}V_{tb}$ and $\rho_{tt}V_{tb}$, respectively,
and $cg \to bH^+$ is {\it not} CKM-suppressed.
Although this fact can be read off from the $H^+$ coupling
in Eq.~(\ref{eq:Yuk}), it is not quite ``apparent'',
and was realized after a study\cite{Hou:2019uxa} of 
$H^+$ effect in $B \to \mu\bar\nu$.
We turn to the $cg \to bH^+ \to bt\bar b$ process in this section.
 The process was noted by others,~\cite{Iguro:2017ysu,Gori:2017tvg,Nierste:2019fbx}
 but were either mentioned without collider study, or without sufficient detail.
 For example, the $gg \to \bar cbH^+$ process was
discussed in Ref.~\refcite{Gori:2017tvg}, but Fig.~\ref{feyndiag}[left] 
was not explicitly mentioned, and in absence of a detailed collider study, 
the promise was not quite elucidated.

\subsubsection{Constraints on Higgs and Flavor Parameters}

We express~\cite{Davidson:2005cw,Hou:2017hiw} 
the quartic couplings $\eta_1$, $\eta_{3{\rm -}6}$ in terms of 
$\mu_{22}$, $m_{h,\, H,\, A,\, H^+}$
 (which are all normalized to $v$) and $\cos\gamma$, 
plus $\eta_2$, $\eta_7$ that do not enter Higgs masses.
As Yukawa couplings of $H^+$ do not depend on $c_\gamma$,
which is known to be small,
we set  $c_\gamma = 0$ (thus, e.g. $t \to ch$ does not constrain $\rho_{tc}$) 
while fixing $m_h \cong $ 125 GeV.
Thus,~\cite{Hou:2017hiw} $\eta_6 = 0$ and $\eta_1 = m_h^2/v^2$,
which simplifies the study considerably.
In the Higgs basis, we identify $\eta_{1-7}$ with the input parameters 
$\Lambda_{1-7}$ to 2HDMC,\cite{Eriksson:2009ws} which we utilize 
to check they satisfy positivity, perturbativity and tree-level unitarity.

\begin{figure}[t]
\center
\includegraphics[width=0.45 \textwidth]{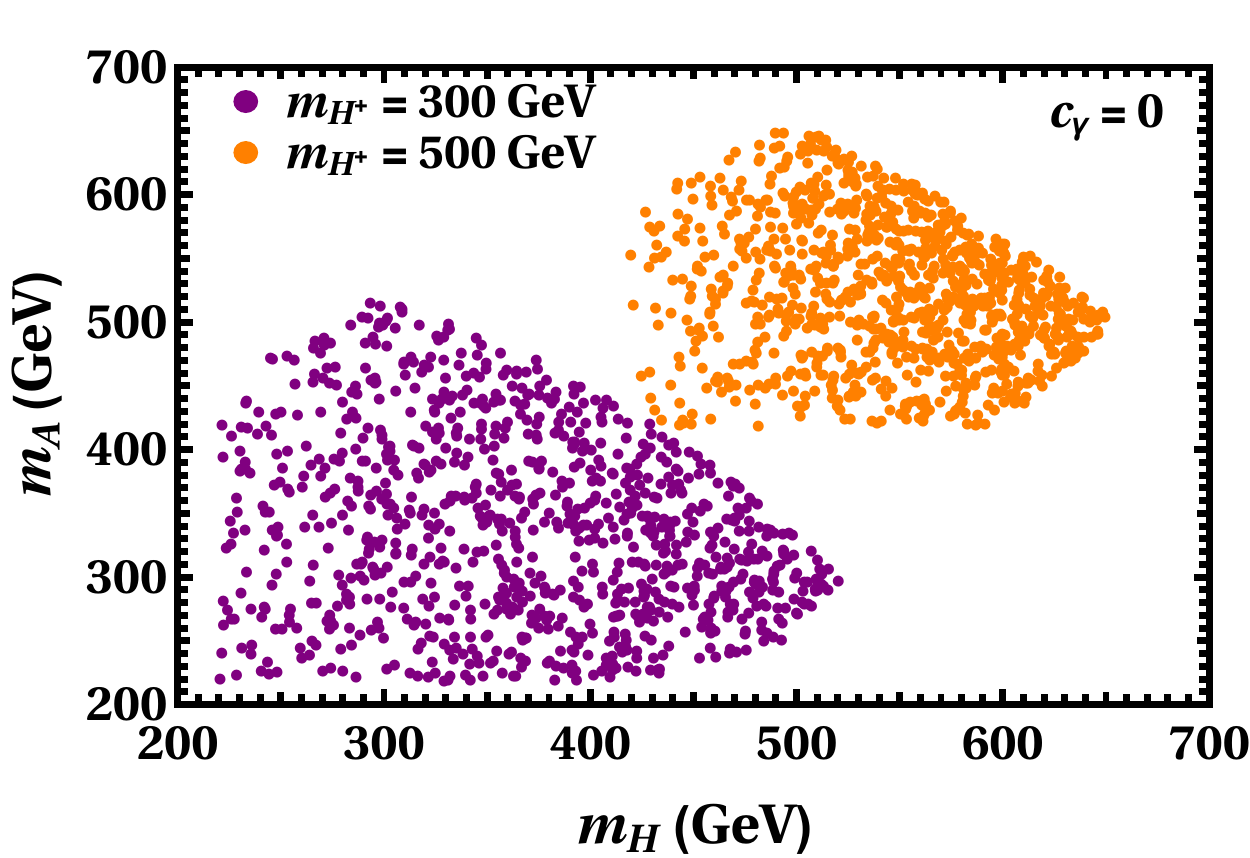}
\caption{
Scan points in $m_H$--$m_A$ plane for $c_\gamma = 0$ that pass 
positivity, perturbativity, unitarity and $T$ parameter constraints.\cite{Ghosh:2019exx} 
See text for details.
}
\label{massscan}
\end{figure}

For fixed $m_{H^+} = 300$ and 500 GeV, 
we randomly generate the parameters in the ranges 
{$\eta_{2-5,\,7} \leq 3$ (except $\eta_2 > 0$ by positivity),} 
$\mu_{22} \in [0, 1]$  TeV,
and $m_{A,\, H} \in [m_{H^+}-m_W, 650$~GeV]
to forbid $H^+ \to AW^+, HW^+$, again to simplify. 
The generated parameters are passed to 2HDMC for scanning, 
where we follow the scanning procedure discussed in Ref.~\refcite{Hou:2019qqi}.
We further impose the electroweak oblique parameter constraints,\cite{T-param}
which restricts\cite{Froggatt:1991qw,Haber:2015pua} 
the scalar masses hence the $\eta_i$s. 
{Scan points satisfying these constraints}\cite{Ghosh:2019exx}
are plotted in Fig.~\ref{massscan} in the $m_H$--$m_A$ plane
 for $m_{H^+} = 300$, 500 GeV,
which illustrates that finite parameter space exist.
We choose a benchmark for each $m_{H^+}$ value  
and list the parameters in Table~\ref{bench}.

\begin{table}[t]
\tbl{
Benchmark point BP1 (BP2) for $m_{H^+} =$ 300 (500) GeV, 
with $\eta_6 = 0$ hence $\eta_1 \cong 0.258$ (alignment limit). 
Higgs masses are in GeV.}
{
\begin{tabular}{c c c c c  c  c c  c c} 
 \toprule
 & 
 $\eta_2$ & $\eta_3$ & $\eta_4$ & $\eta_5$ 
 & $\eta_7$
 & \,${\mu_{22}^2/v^2}$\, &  $m_{H^+}$  &  $m_A$  &  $m_H$  \\
\hline
\,BP1\, & 
  1.40 & 0.62 & 0.53 & \,\;\ 1.06 & 
 {$-0.79$} & 1.18 & 300 & 272 & 372 \\
\,BP2\, & 
 0.71 & 0.69 & 1.52 & {$-0.93$} & 
 {\,\;\ 0.24} & 3.78 & 500  & 569   & 517 \\
%
 \botrule
\end{tabular}
\label{bench}}
\end{table}

\vskip0.15cm
\noindent\underline{\it Flavor Constraints}
\vskip0.05cm

\begin{figure*}[t]
\center
\includegraphics[width=0.41 \textwidth]{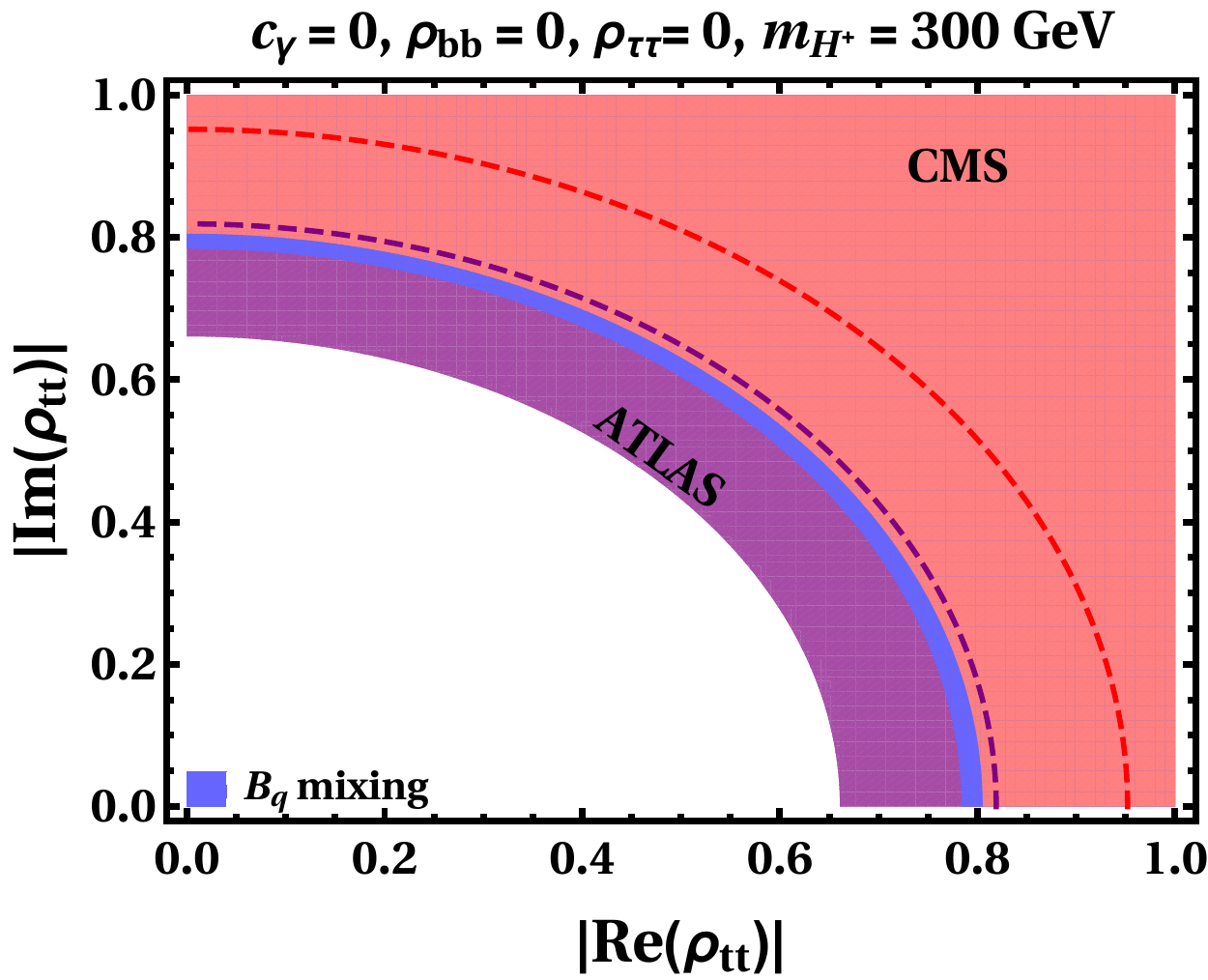}
\includegraphics[width=0.41 \textwidth]{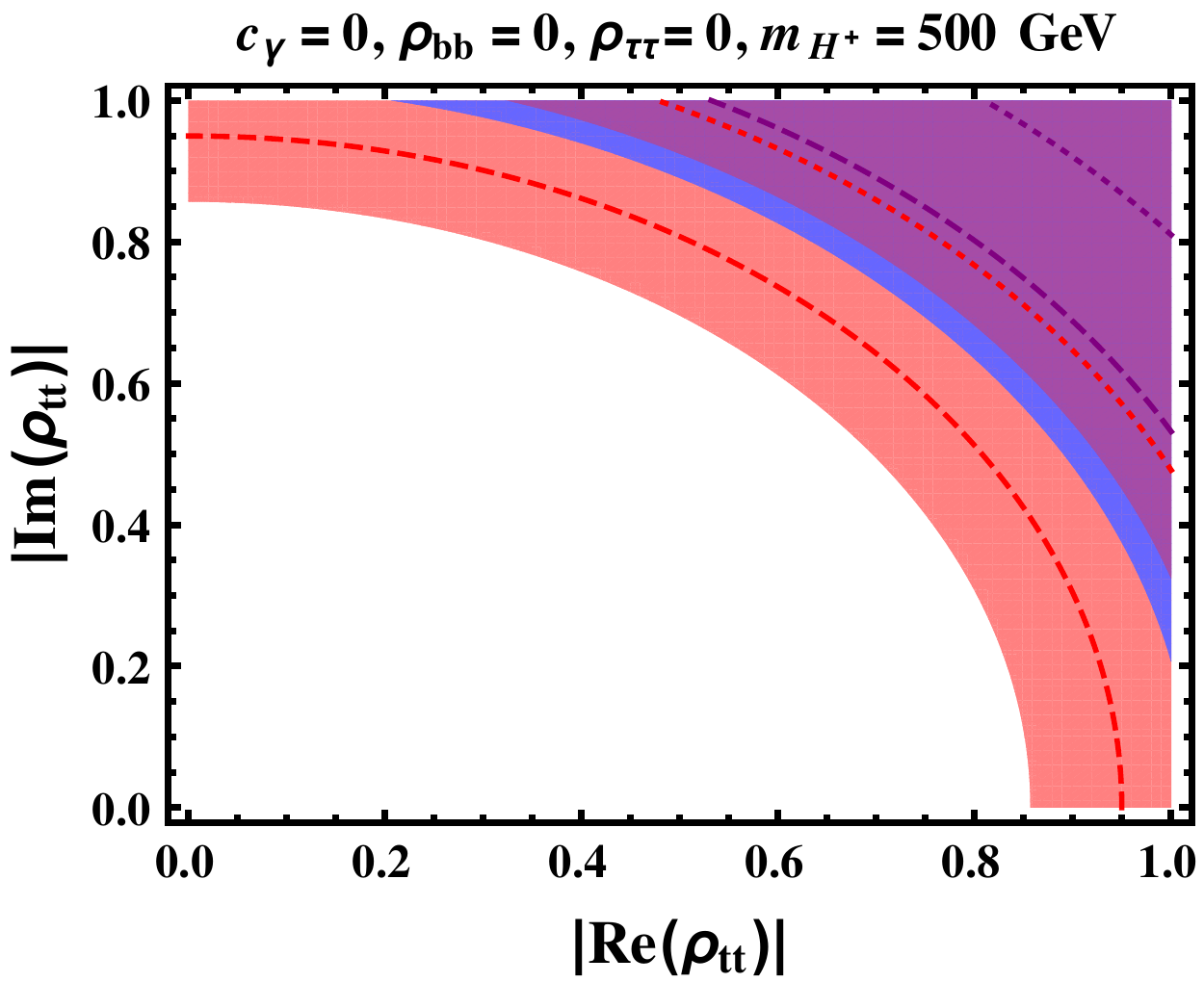}
\caption{
Constraint from $B_q$ mixings on $\rho_{tt}$
 (blue shaded region with area to upper right excluded)
 assuming all other $\rho_{ij}=0$.  
The excluded regions from $bg \to \bar t (b) H^+ \to \bar t (b) t \bar b$ searches 
by ATLAS~\cite{Aaboud:2018cwk} and CMS~\cite{Sirunyan:2019arl} 
for $\rho_{tc} = 0$ are overlaid (purple and red shaded),
which is weakened for $\rho_{tc}=0.4$ (dash) 
{and 0.8 (dots).\cite{Ghosh:2019exx} See text for details}.
}
\label{rhottconst}
\end{figure*}

With large parameter space for the Higgs sector demonstrated,
we turn to flavor and collider constraints on $\rho_{tc}$ and $\rho_{tt}$.
Flavor constraints 
are not very strong~\cite{Chen:2013qta,Altunkaynak:2015twa}. 
For $m_{H^+} \lesssim 500$ GeV, $B_q$ 
mixings ($q$ = $d$, $s$) provide the relevant constraint. 
In particular, and as mentioned before,
$\rho_{ct}$ must be turned off~\cite{Altunkaynak:2015twa}
because of a CKM-enhanced effect in $B_q$ mixings.
Assuming all $\rho_{ij}$ vanish except $\rho_{tt}$,
we define $M^q_{12}/M^{q}_{12}|^{\rm{SM}}= C_{B_q}$, with phase negligible.
Allowing $2\sigma$ error on 
$C_{B_d} = 1.05\pm 0.11$ and $C_{B_s} = 1.11\pm0.09$ 
from summer 2018 UTfit~\cite{utfitrse}, we give the 
blue shaded exclusion region in Fig.~\ref{rhottconst}, 
which extends to upper-right and the left (right) panel is for BP1 (BP2).
 {The constraint from $H^+$ effects via charm loops~\cite{Crivellin:2013wna} 
 is more forgiving.
 }

$B \to X_s\gamma$ puts a strong constraint on $m_{H^+}$ in 2HDM II, 
but weakens for g2HDM {due to extra Yukawa couplings}.
An $m_t/m_b$ enhancement factor actually constrains $\rho_{bb}$ 
more strongly~\cite{Altunkaynak:2015twa} than $\rho_{tt}$. 
Taking $\rho_{bb}$ as small, 
the constraint on $\rho_{tt}$ from $B \to X_s \gamma$ 
falls outside the range of Fig.~\ref{rhottconst},
while the $B \to X_s\gamma$ constraint on $\rho_{tc}$ via charm loop
is weaker than $B_q$ mixing~\cite{Altunkaynak:2015twa}.

Overall, because many parameters enter, we view the 
true constraints from $B \to X_s\gamma$ on $H^+$ parameters, 
e.g. $m_{H^+}$, as still an open issue.

\vskip0.15cm
\noindent\underline{\it Collider Constraints}
\vskip0.05cm

For collider constraints, we again set all $\rho_{ij} = 0$ 
 except $\rho_{tt}$ and $\rho_{tc}$ for simplicity. 
For finite $\rho_{tt}$, one can have\cite{Gunion:1986pe, DiazCruz:1992gg, Moretti:1999bw, Miller:1999bm, Arhrib:2018ewj, Arhrib:2019ykh, Cen:2018okf, Sanyal:2019xcp, Coleppa:2019cul} 
$\bar bg \to \bar t (b) H^+$\,(our list is not exhaustive),
followed by $H^+\to t \bar b$. 
Searches at $13$\,TeV provide model independent bounds on
$\sigma(pp\to \bar t {(b)} H^+) \,  \mathcal{B}(H^+  \to t \bar b)$ 
for $m_{H^+}=200$\,GeV to  2\,(3)\,TeV
 for ATLAS~\cite{Aaboud:2018cwk} (CMS~\cite{Sirunyan:2019arl}). 
Using MadGraph5\_aMC@NLO as before with 
default PDF of NN23LO1 and effective model implemented in FeynRules, 
we calculate $\sigma(pp\to \bar t {(b)} H^+)\, (H^+  \to t \bar b)$ 
at LO for a reference $|\rho_{tt}|$, then rescale by 
$|\rho_{tt}|^2\,\mathcal{B}(H^+ \to t \bar b)$ to get the upper limits. 
For $m_{H^+} = 300$, 500 GeV and 
with $\rho_{tc}=0$ (hence $\mathcal{B}(H^+ \to t \bar b) \sim 100\%$),
we plot the extracted 
ATLAS (CMS) 95\% C.L. bounds on $\rho_{tt}$ as
the red (purple) shaded regions in Fig.~\ref{rhottconst}.
The ATLAS/CMS limit is more/less stringent than 
$B_q$ mixing for BP1 
while opposite for BP2, 
with the exclusion bands overlaid to illustrate this.
Constraints on $\rho_{tt}$ from\cite{Aaboud:2017hnm,Sirunyan:2019wph} 
$gg\to H/A \to t \bar t$ search are weaker than results shown in Fig.~\ref{rhottconst}.
We will return to the CMS ``excess''\cite{Sirunyan:2019wph} 
at $m_A \sim 400$ GeV later.

The CMS $4t$ search\cite{Sirunyan:2019wxt} based on full Run~2 data
constrains $\rho_{tc}$, as discussed in Sec. 4.1.1,
but now we consider the constraint together with $\rho_{tt}$. 
With both $\rho_{tc}$ and $\rho_{tt}$ finite, 
the $cg\to t H/tA \to t t \bar t$ process~\cite{Kohda:2017fkn} 
can feed the SR12 signal region of the CMS $4t$ search
if all three top quarks decay semileptonically.
As $cg \to tH/tA \to tt\bar t$ barely occurs for BP1 
because of low $m_{A,\, H}$ values, 
this applies only to BP2.
SR12 requires~\cite{Sirunyan:2019wxt} at least three leptons, 
four jets with at least three $b$-tagged, plus $p_T^{\rm miss}$. 
We generate events with PYTHIA~6.4 for showering and hadronization, 
adopt MLM merging\cite{Mangano:2006rw, Alwall:2007fs}
 of matrix element and parton shower, 
then feed into\cite{deFavereau:2013fsa} Delphes~3.4.2 
for CMS-based detector card, 
including $b$-tagging and $c$- and light-jet rejection.
We find $\rho_{tt} \gtrsim 1$ is excluded if $\rho_{tc} \sim 0.8$
for BP2.
However, finite $\rho_{tc}$ induces $H^+ \to c\bar b$ decay, 
which would dilute ${\cal B}(H^+ \to t\bar b)$ and 
soften the $bg \to \bar t (b) H^+$ constraint. 
This is illustrated by the dash (dot) curves
in Fig.~\ref{rhottconst}(right) for $\rho_{tc} = 0.4\, (0.8)$.
%

As already discussed in Sec.~4.1.1, 
the $cg \to tH/tA \to tt\bar c$ process~\cite{Kohda:2017fkn} 
can feed the CRW control region of CMS 4$t$ study 
when both tops decay semileptonically. 
Following Refs.~\refcite{Hou:2018zmg}, \refcite{Hou:2019gpn}, 
we find $\rho_{tc}\gtrsim 0.4$ is excluded for BP1, 
which is stronger than the $B_q$ mixing bound, and
with little dependence on $\rho_{tt}$.
For BP2, CRW gives comparable limit as SR12.
Thus, we give in Fig.~\ref{rhottconst}(left) 
the softened $bg \to \bar t (b) H^+$ constraint 
only for $\rho_{tc} = 0.4$.
Note that analogous searches with similar signature
usually involve stronger cuts and
therefore do not give relevant constraints.

\subsubsection{Collider Signature for $cg \to bH^+ \to bt\bar b$}

We now show that the $cg \to bH^+ \to bt\bar b$ process,
or $pp \to b H^+ + X \to b t \bar b +X$, is quite promising.\cite{Ghosh:2019exx}
For illustration, we {\it conservatively} take 
$|\rho_{tc}|=0.4$, $|\rho_{tt}|= 0.6$ for both BPs. 
%
{Receiving no CKM suppression, the approximate} 
$H^+ \to c\bar b,\; t\bar b$ branching ratios 
are 50\%, 50\% for BP1, and 36\%, 64\% for BP2.
%
%
Assuming $t\to b \ell \nu_\ell$ ($\ell =e,\,\mu$),
the signature is one charged lepton, $p^{\rm miss}_T$, and three $b$-jets. 
%
%
{
The inclusive final state will receive contributions 
from $bg\to \bar c H^+\to \bar  c t \bar b$ 
and $bg\to \bar tH^+\to \bar  t c \bar b$, as well as 
$\rho_{tt}$-induced $bg\to \bar tH^+\to \bar  t t \bar b$. 
Furthermore, the $3b1\ell$ signature will receive mild contributions 
from $gg \to c \bar b H^+$ and $gg \to t \bar b H^+$ processes. 
To include such contributions, we have generated 
$cg\to bH^+$ events 
with up to two additional partons, 
and $bg\to \bar tH^+$ events with one additional parton.
}
%
The $c\bar b \to H^+ \to c \bar b, t \bar b$ processes\footnote{
 The inclusive $\sigma(pp \to H^+)$ with up to two extra jets 
 in the five flavor scheme for BP1 (BP2) at 14\,TeV is 50.8 (6.5) pb 
 using MadGraph5\_aMC with NN23LO1 PDF set and default run card.
} 
suffer from QCD and top backgrounds for hadronic decay of top, while 
$c\bar b \to H^+ \to t\bar b$ with semileptonic decay of top is accounted in signal.
%
The dominant backgrounds for $cg \to bH^+$  arise from 
$t\bar t+$jets, $t$- and $s$-channel single-top ($tj$), $Wt+$jets, 
with subdominant backgrounds from $t\bar t h$ and $t\bar t Z$. 
Minor contributions from Drell-Yan and $W+$jets, $4t$, $t\bar t W$, $tWh$ 
are {combined} under ``other''.

Signal and background samples are generated at LO 
for $14$ TeV as before by MadGraph, interfaced with PYTHIA 
and fed into Delphes for fast detector simulation adopting 
default ATLAS-based detector card. 
The LO $t\bar t +$jets background is normalized to
NNLO by a factor~\cite{twikittbar} $1.84$, and 
factors of~\cite{twikisingtop} 1.2 and 1.47 for $t$- and $s$-channel single-top. 
The LO $Wt+$jets background is normalized to NLO 
by a factor~\cite{Kidonakis:2010ux} 1.35, 
whereas the subdominant $t\bar t h$, $t\bar t Z$ receive 
factors of 1.27~\cite{twikittbarh}, 1.56~\cite{Campbell:2013yla}.
The DY+jets background is normalized to NNLO 
by a factor~\cite{Li:2012wna} 1.27. 
Finally, the $4t$ and $t\bar t W^-$ ($t\bar t W^+$) cross sections at LO 
are adjusted to NLO by factors of  2.04\cite{Alwall:2014hca}
and\cite{Campbell:2012dh} 1.35 (1.27).
The $tWh$ and $W+$jets backgrounds are kept at LO.
Correction factors for other charge conjugate processes are assumed to be the same, 
and the signal cross sections are kept at LO.

Events are selected with one lepton with $p_T^\ell > 30$ GeV, 
at least three jets with three $b$-tagged and with $p_T^b >20$ GeV, 
$|\eta|<2.5$ for lepton and the $b$-jets, and $E^{\rm miss}_T  >35$ GeV. 
Jets are reconstructed by anti-$k_t$ algorithm using $R=0.6$. 
Whether between $b$-jets or the lepton, 
we require their $\Delta R > 0.4$. 
The sum of the lepton and three leading $b$-jet transverse momenta $H_T$ 
should be $>350$ (400) GeV for BP1 (BP2). 
%
%
The selection cuts for 
$H_T$, $p_T$, $E^{\rm miss}_T$, etc. are not optimized. 
The total background~B$_{\rm tot}$ (and its various components)
and signal Sig cross sections after selection cuts 
are given in Table~\ref{bkgcomptbhpm}.

\begin{table}[t]
\tbl{
Background and signal (Sig, for $\rho_{tc} = 0.4$, $\rho_{tt} = 0.6$) 
cross sections (in fb) at $14$ TeV after selection cuts.}
{
\begin{tabular}{c c c c c  c  c c  c} 
 \toprule
  & $t\bar tj$s & $tj$ & $Wtj$s & $t\bar t h$ & $t\bar t Z$ & other
  & B$_{\rm tot}$ & Sig  \\
\hline
 \,BP1\, & \,1546\, & \,42\, & 27 & \,4.2\, & 1.5 & 3.1 & \,1627\, & \,11.4\,  \\
 \,BP2\, & \, 1000 \, & 27 & 16 & 2.9 & 1.2 & 1.9 &\  1049\, & \,9.3\,   \\
 \botrule
\end{tabular}
\label{bkgcomptbhpm}
}
\end{table}

We estimate the statistical significance from Table~\ref{bkgcomptbhpm} 
using Eq.~(\ref{poissn}).
For 137, 300 and 600 fb$^{-1}$, the significance for $cg \to bH^+$ is at
$\sim 3.3\sigma$, $4.9\sigma$, $6.9\sigma$ 
($\sim 3.4\sigma$, $5.0\sigma$, $7.1\sigma$) for BP1 (BP2).
{{Reanalyzing} 
for 13 TeV 
at $137$ fb$^{-1}$, we find similar significance}
{\it per} experiment.
Thus, full Run~2 data could already show evidence, 
and combining ATLAS and CMS data is encouraged.
Discovery is possible for Run~2+3 and beyond. 
Note that significance can still be high at higher masses 
for larger $\rho_{tc}$, $\rho_{tt}$, but the decoupling $\mu_{22}^2$ 
would generally become larger,\cite{Hou:2017hiw} 
as can be peeked from $\mu_{22}^2/v^2 \simeq 3.78$ for BP2 
in Table~\ref{bench}, which would start to damp the EWBG motivation.
The $cg \to bH^+$ process, however, can certainly be pursued 
for heavier $m_{H^+}$ at higher luminosities.

Our BPs here involve both $\rho_{tc}$ and $\rho_{tt}$,
and can feed the SS$2\ell$ signature.
Following the same analysis of Refs.~\refcite{Kohda:2017fkn}
and \refcite{Hou:2018zmg}, we find 
BP1 may have $\sim3.5\sigma$ significance with full Run 2 data
since finite $\rho_{tt}$ plays no effect. 
But for BP2, the significance is below $\sim1\sigma$ due to 
dilution from $A/H\to t\bar t$ decay and falling parton luminosity. 
Single-top studies may contain $cg \to bH^+$ events.
For $\rho_{tc} = 0.4$ and  $\rho_{tt} = 0.6$, 
we find the combined cross sections for $pp\to H^{+}[t\bar b] j$, 
$H^{+}[c\bar b] t$  
can contribute 15.2 (2.9)\,pb for BP1 (BP2),
which is within the $2\sigma$ error of current $t$-channel 
single-top~\cite{Aaboud:2016ymp,Sirunyan:2018rlu} measurements.
The situation is similar for Run 1 with $s$-channel single-top measurements.
We have not included uncertainties from scale dependence and PDF, where the latter is 
sizable for processes initiated by heavy quarks.~\cite{Buza:1996wv,Maltoni:2012pa} 
Using signal cross sections at LO can also bring in some uncertainties,
e.g. higher order corrections\cite{Kidonakis:2010ux,Plehn:2002vy,Berger:2003sm}
to $\sigma(bg\to t H^+)$ may amount to $30$--$40\%$ for $m_{H^+} \sim 300$--500\,GeV. 
A detailed study of such uncertainties is left for the future,
and is part of the reason why we adopt 
conservative $\rho_{tc}$, $\rho_{tt}$ values.

\section{Discussion and Miscellany}

With $\rho_{tt} = {\cal O}(\lambda_t) \sim 1$,
the heavy Higgs bosons $H$ and $A$ can be produced 
via gluon-gluon fusion (ggF), and $gg \to H/A \to t\bar t$ 
can interfere with the large, QCD-induced $gg \to t\bar t$ amplitude, which 
distorts the Breit-Wigner peak into a peak-dip structure,\cite{Carena:2016npr} 
making experimental search more challenging.
A first study by ATLAS with 8 TeV data starting from 500~GeV 
found\cite{Aaboud:2017hnm} no significant deviation.
Based on 2016 data, CMS reported\cite{Sirunyan:2019wph} more 
recently an intriguing signal-like deviation in $gg \to A \to t\bar t$, 
compatible with an $A$ with mass $\sim 400$\,GeV with 
global significance of $1.9\sigma$ (and $3.5\sigma$ local).
It was shown\cite{Hou:2019gpn} that $\rho_{tt} = {\cal O}(1)$ 
can account for this possible ``excess'', but $\rho_{tc} = {\cal O}(1)$ 
is called upon to reign in ${\cal B}(A \to t\bar t)$.
It was further pointed out that,\cite{Hou:2019gpn} for purely 
imaginary $\rho_{tt}$ --- which can robustly drive\cite{Fuyuto:2017ewj} EWBG --- 
an $H$ would mimic\cite{Hou:2018uvr} an $A$ in production and decay, 
and CMS could well be peeking at the first of the exotic bosons, 
$m_H \sim 400$\,GeV, with heavier, degenerate 
$m_A \simeq m_{H^+} \gtrsim 550$\,GeV satisfying\cite{Hou:2017hiw} 
custodial symmetry, hence more easily fulfill oblique parameter constraints.\footnote{
 Having $m_A = 400$\,GeV while $m_H \simeq m_{H^+} \gtrsim$\,550\,GeV
 would correspond to ``twisted'' custodial symmetry\cite{Gerard:2007kn}
 with more restricted parameter space.
} 
This fits nicely in the range of Eq.~(\ref{eq:mrange}),
and having $A$--$H^+$ heavier\footnote{
 We remark that, based on $\sim 36$\,fb$^{-1}$, a relatively stringent
 CMS bound\cite{Sirunyan:2020hwv} on $m_{H^+}$ from
 $pp \to \bar t H^+ +X$ associated production followed by
 $H^+ \to t\bar b$ would push $m_{H^+}$ beyond 650\,GeV.
 Full Run~2 data, including that from ATLAS, can tell
 whether this is a downward fluctuation.
} 
also fits the single-state analysis of CMS.\cite{Sirunyan:2019wph}
We eagerly await the unveiling of the full Run\,2 data from both CMS and ATLAS. 

But perhaps the excess would disappear with full Run~2 analysis,
or some analysis difficulty may be encountered.
There is then the $gg \to H/A \to t\bar c$ process,\cite{Altunkaynak:2015twa}
mediated by $\rho_{tt}$ in production and $\rho_{tc}$ in decay.
This could be promising, as it does not suffer from interference.
However, there is worry\cite{Kohda:2017fkn} that 
$tj$ ($j$ stands for a jet) mass resolution\cite{Sirunyan:2017yta} 
could wash the resonance away.
Or one could pursue $H/A \to \tau\mu$ in the final state,
which was shown\cite{Hou:2019grj} to hold promise at the HL-LHC, 
but would likely suffer from suppressed ${\cal B}(H/A \to \tau\mu)$,
as $\rho_{\tau\mu} = {\cal O}(\lambda_\tau) \sim 0.01$ 
would be our best guess.\cite{Hou:2020itz}
In principle, $H/A \to \tau\tau$ can also be searched for,
but there is no ``large $\tan\beta$'' enhancement mechanism as in 2HDM\,II, 
where again our best guess would be $\rho_{\tau\tau} = {\cal O}(\lambda_\tau)$
in g2HDM, that the second diagonal $\tau$ Yukawa coupling should go with
the strength of $\lambda_\tau$ by the mass-mixing hierarchy argument.
We note the prowess of ATLAS\cite{Aad:2020zxo} and CMS\cite{Sirunyan:2019shc} 
with simple final states containing $\tau$s.

This brings us back to our main proposed processes of
$cg \to tH/A \to tt\bar c,\, tt\bar t$ and $cg \to bH^+ \to bt\bar b$,
but turning to the $\rho_{tu}$-induced process.
Although $\rho_{tu} < \rho_{tc}$ is expected by way of mass-mixing argument,
this is not based on direct experimental knowledge.
In Sec.~4.1.1., we followed Ref.~\refcite{Hou:2020ciy}
to study the effect of CRW of CMS 4$t$ analysis 
and CRttW2$\ell$ of ATLAS 4$t$ analysis on $\rho_{tu}$,
and plotted the results in Fig.~\ref{exclu}[right].
The bounds from $4t$ analyses on $\rho_{tu}$ are indeed more stringent
than on $\rho_{tc}$ of Fig.~\ref{exclu}[left], but by far not as stringent
by a $\sqrt{m_u/m_c}$ factor as implied roughly by the mass-mixing hierarchy.
Furthermore, these bounds were studied keeping\cite{Hou:2020ciy}
$\rho_{tu} \neq 0$ but setting all other $\rho_{ij} = 0$,
with exclusion limits and discovery reach also 
plotted in Fig.~\ref{exclu}[right].
If $\rho_{tu}$ and $\rho_{tc}$ are both kept,\cite{Hou:2020ciy} 
given that there are no good tools for separating $c$- and light $q$-jets, 
it would not be easy to gain on extracting information.
We note, therefore, that the product $\rho_{tu}\rho_{\tau\mu}$
would be probed by\cite{Hou:2019uxa} the $B \to \mu\nu$ process at Belle\,II. 
If any deviation\cite{Chang:2017wpl} of ${\cal B}(B \to \mu\nu)/{\cal B}(B \to \tau\nu)$ 
from SM expectation is observed, it would imply that both $\rho_{tu}$ {\it and} $\rho_{\tau\mu}$ are nonvanishing, and more sophisticated analysis 
for having $\rho_{tu}$ and $\rho_{tc}$ both finite need to be developed,
where Ref.~\refcite{Hou:2020ciy} is only a start. 
The $\rho_{\tau\mu}$ coupling can also be probed via the $\tau \to \mu\gamma$ 
process\cite{Hou:2020tgl, Hou:2020itz} at Belle\,II with the help of 
$\rho_{tt}$ via the two-loop mechanism.\cite{Chang:1993kw}
Thus, there is much synergies with the flavor frontier to look forward to,
and with refined analyses at the (HL-)LHC, much progress can be anticipated.

Finally, let us comment on the implications of finite $c_\gamma$.
For exotic Higgs in the mass range of Eq.~(\ref{eq:mrange}),
in general one would expect $h$-$H$ mixing, i.e. $c_\gamma$, to be finite.
However, compared with the $\cos(\beta-\alpha)$ fit to Higgs property 
measurements\cite{Sirunyan:2018koj, Aad:2019mbh} at the LHC 
in 2HDM\,II (which is implied by SUSY), 
the many more parameters that can enter for production 
and decay of the $h$ boson, as we have illustrated only partially,
implies that a similar fit to $c_\gamma$ (the equivalent to $\cos(\beta-\alpha)$)
seems nowhere in sight. We can only argue that $c_\gamma$ has to be
small to have $h$ resembling the Higgs boson of SM so well,
but its value is not known.

With $c_\gamma$ small but nonzero, what could be the implications?
The $H$ and $A$ (and $H^+$) would become mildly related to 
electroweak symmetry breaking, e.g. coupling to $W$ and $Z$ bosons.
We mention here only a few collider possibilities.
The first would be resonant di-Higgs production via the
$cg \to tH \to thh$ process, where a finite $c_\gamma$ induces
an effective $\lambda_{Hhh}$ coupling.\cite{Hou:2019qqi} 
This can be searched for at the LHC via $pp \to tH + X \to thh +X$,
and non-negligible discovery potential is found\cite{Hou:2019qqi} 
at the LHC for $m_H \lesssim 350$\,GeV and $\rho_{tc} \gtrsim 0.5$.
A second process would be $cg \to tA \to tZH$,\cite{Hou:2019mve} 
where sufficient $m_A$--$m_H$ splitting is required to allow for
$A \to ZH$ decay. It further requires $\rho_{tt}$ to be small,
otherwise $A \to t\bar t$ would in general be too strong.
Under these conditions, it is found\cite{Hou:2019mve} that
discovery is possible at the HL-LHC for $m_A \sim 400$ GeV,
hence $m_H \lesssim 300$\,GeV, but $tZh$ production is not promising.
A third example, discussed recently,\cite{HMP2012} targets the conclusive
investigation of the $\rho_{tc}$ mechanism\cite{Fuyuto:2017ewj} for EWBG 
(in the case that $\rho_{tt}$ turns out small),
which can more easily\cite{Fuyuto:2019svr} survive the eEDM constraint 
as $\rho_{tc}$ does not enter the Barr-Zee two-loop mechanism.
The $\rho_{tc}$ mechanism requires $\rho_{tc} \gtrsim 0.5$.
The proposal\cite{HMP2012} is to utilize $cg \to bH^+ \to bW^+h$ for 
$c_\gamma \gtrsim 0.1$ to efficiently push the $\rho_{tc}$ bound
(or discovery!) down to 0.2 or below at the HL-LHC. 
For small $c_\gamma \lesssim 0.12$, one counts on the $cg \to tH/A \to tt\bar c$ process,
which depends rather weakly on $c_\gamma$, to push the exclusion
(or discovery!)  down to 0.25 or below.
Combined, this two-prong search can draw a conclusion 
on the $\rho_{tc}$ mechanism.

\section{Summary and Prospect}

In the past few years, we have verified the Yukawa couplings of $t$, $b$ quarks
and $\tau$, $\mu$ charged leptons, making extra Yukawa couplings involving 
a second Higgs doublet plausible and attractive. 
This general 2HDM with extra Yukawa couplings point to 
extra Higgs bosons $H$, $A$ and $H^+$ in the sub-TeV range, which is
based on naturalness of the second set of dimension-4 couplings
unique to the Higgs sector: Higgs quartic self-couplings.
The exotic bosons are well hidden so far by fermion mass-mixing hierarchy
and alignment, the smallness of $h$--$H$ mixing angle $c_\gamma$.

The search for $t \to ch$ (and concurrently for $h \to \tau\mu$) is ongoing. 
If it does not emerge with full Run~2 data, it would constrain
the product of $\rho_{tc}\,c_\gamma$. But given that $c_\gamma$
is small, sizable $\rho_{tc}$ would still be allowed.
In this brief review, we advocate the search for
$cg \to tH/A \to tt\bar c,\,tt\bar t$ and $cg \to bH^+ \to bt\bar b$ at the LHC,
where production, unhampered by small $c_\gamma$, depends on sizable $\rho_{tc}$, 
while the $tt\bar t$ and $bt\bar b$ final states require finite $\rho_{tt}$
for $H/A \to t\bar t$ and $H^+ \to t\bar b$ decays.
It would be interesting, therefore, to follow up with full Run~2 data on the 
$m_A \sim 400$\,GeV excess seen by CMS in $t\bar t$ resonance search.
Both extra top Yukawa couplings $\rho_{tt}$ and $\rho_{tc}$ 
can drive electroweak baryogenesis, providing further impetus for search. 
LHC Run~2 data may already be promising,
and Run 2+3 data would be even more revealing,
while HL-LHC may hold the ultimate promise for discovery,
which would open up a new chapter at the intersection of Higgs and flavor physics.

\section*{Acknowledgments}

WSH is supported by MOST 109-2112-M-002-015-MY3 of Taiwan
and NTU 109L104019. TM is supported by a Postdoctoral Research 
Fellowship from Alexander von Humboldt Foundation.

\end{document}